\documentclass{myaa}
\usepackage{epsf}
\usepackage{psfig}
\usepackage{graphics}

\def\gal{ESO~342--G017}
\def\kms{km\,s$^{-1}$}
\def\Msqarc{\,mag/sq~arcsec}
\def\SeX{{\tt SeXtractor}}
\def\mid{{\tt MIDAS}}
\def\obj{{\tt OBJECTS}}

\begin{document}


\title{Detection of a Thick Disk in the edge-on Low Surface Brightness
	Galaxy ESO~342--G017$^{\star}$}

\subtitle{I. VLT Photometry in $V$ and $R$ Bands}

\author{Mark J. Neeser \inst{1,2}
	\and Penny D. Sackett \inst{1}
	\and Guido De Marchi \inst{3}
	\and Francesco Paresce \inst{4}}

\institute{  Kapteyn Astronomical Institute,
                Postbus 800,
                9700 AV Groningen,
                The Netherlands
\and
		Universit\"ats--Sternwarte M\"unchen,
		Scheinerstr. 1,
		D--81679 M\"unchen,
		Germany
\and
		European Space Agency,
		Research and Science Support Department,
		3700 San Martin Drive,
		Baltimore, MD 21218, 
		USA
\and
		European Southern Observatory,
                Karl-Schwarzschild-Str.\,2,
                D--85748 Garching bei M\"unchen,
                Germany
             }

\offprints{M.\,J.\, Neeser (neeser@usm.uni-muenchen.de)\\
$^\star$Based on observations collected at the European Southern Observatory, 
Paranal, Chile (VLT-UT1 Science Verification Program)}

\date{Received: 30 September 2001; Accepted: 11 December 2001}


\titlerunning{\ \ Thick Disk in LSB Spiral ESO~342--G017}

\authorrunning{Neeser et al.}

\abstract { We report the detection of a thick disk in the edge-on,
low surface brightness (LSB), late-type spiral \gal, based on
ultra-deep images in the $V$ and $R$ bands obtained
with the VLT Test Camera during Science Verification on UT1.
All steps in the reduction procedure are fully described,
which, together with an extensive analysis of systematic and
statistic uncertainties, has resulted in surface brightness
photometry that is reliable for the detection of faint
extended structure to a level of $V = 27.5~$ and
$R = 28.5~$\Msqarc.  The faint light apparent in these
deep images is well-modeled by a thick exponential disk
with an intrinsic scale height about 2.5 times that of the
thin disk, and a comparable or somewhat larger scale length.
Deprojection including the effects of inclination and
convolution with the PSF allow us to estimate that the
thick disk contributes 20-40\% of the total (old) stellar
disk luminosity of \gal.
To our knowledge, this is the first detection of a thick disk in
an LSB galaxy, which are generally thought to be rather unevolved
compared to higher surface brightness galaxies.
}

\maketitle

%

\section{Introduction}\label{intro}

Outside our own Galaxy, most of what we know about the structure,
evolution and dynamics of stellar populations, and their connection to
dark matter, is deduced from high surface brightness features: bars,
bulges, and thin disks.  Fainter surface brightness components such as
stellar halos, thick disks, and globular clusters probe galactic
potentials differently, in both time and space owing to their larger
age and extent.  The formation mechanisms of these faint tracers are
still a matter of some controversy; suggestions range from early
protogalactic collapse, secular processes such as heating from
molecular clouds, black holes and spiral structure, through to later
stochastic processes such as accretion (see recent reviews by
\cite{buser00}; \cite{bhf00}; and references therein).  These scenarios
predict different kinematical, morphological and chemical
characteristics, but too few systems have been sufficiently well
studied to constrain the models.  Due to the difficulty in detecting low
surface brightness features reliably in external galaxies, the
important complementary information they contain has only begun to be
tapped.

In the Milky Way, faint disk and halo components can be separated on
the basis of their kinematics and morphology, and -- to a certain
extent -- metallicity, because individual stars can be resolved.  The
Galactic stellar halo of field stars and the globular cluster systems
have volume densities that decrease with galactocentric radius $r$
roughly as $\rho (r) \propto r^{-3.0}$ or $r^{-3.5}$ (\cite{harris79};
\cite{saha85}; \cite{zinn85}), similar to results for halo populations
in large spirals like M31 (\cite{racine91}; \cite{reitzel98}) and
NGC~4565 (\cite{fleming95}).
Giant ellipticals and superluminous CD galaxies, on the other hand,
which are thought to be the product of many mergers, have halo
luminosities and globular cluster systems that fall less steeply,
roughly as $\rho (r) \propto r^{-2.3}$ (\cite{harris86};
\cite{bridges91}; \cite{harris95}; \cite{graham96}).  The total mass,
the bulk of which is believed to be contained in dark matter halos, is
inferred from kinematical studies to have volume densities that decline
as $\rho (r) \sim r^{-2}$ beyond a few disk scale lengths (see
\cite{sackett96} for a review).

Our Galaxy also has a faint thick disk whose density falls
exponentially with increasing height ($z$) above the plane as
$e^{-z/h_{z}^{\rm thick}}$.  Its scale height $h_{z}^{\rm thick} \simeq
1 \pm 0.3$~kpc (\cite{reid93}; \cite{ojha96}; \cite{buser99}) is about
three times larger than that of the much brighter thin disk.   

The scale length of the thick disk is $h_{R}^{\rm thick} \simeq 3 \pm 1.5$~kpc 
(\cite{buser99}), similar to that of the Galactic thin disk.  Despite
this, the thick disk contibutes only 2-9\% of the total local stellar
disk light (\cite{reid93}; \cite{ojha96}; \cite{buser99}),
and perhaps $\sim$13\% of the total disk luminosity of the Milky Way 
(\cite{morrison94}).


For external galaxies, morphology determined through integrated surface
brightness photometry is the only current method to detect and
characterize faint galactic components.  Detections of extended light
that are perhaps indicative of a thick disk component with $h_{z}^{\rm
thick} \simeq 1-2$~kpc have been reported in a few external edge-on
galaxies.  Early detections of extra-planar light in excess of that
associated with a thin exponential disk  were limited to SO
(\cite{burstein79}) and early-type spirals with significant bulges
(\cite{vdkruit81a}; \cite{vdkruit81b}; \cite{wakamatsu84};
\cite{bahcall85}; \cite{shaw89}; \cite{degrijs96}), 
\cite{morrison97}).  leading to the supposition that thick disks were
found in older stellar systems with significant central concentrations
(\cite{vdkruit81a}; \cite{hamabe89}, \cite{degrijs97}).  This
hypothesis is consistent with the lack of a thick luminous component
around the small, Scd spiral NGC~4244 in deep $R$-band observations
reaching to $R = 27.5$\Msqarc\ (\cite{fry99}), and in the 
bulgeless Sd edge-on NGC~7321 (\cite{matthews99}).
On the other hand, observations indicate that there are individual exceptions.
Multiband photometry of the later-type Sc spiral
NGC~6504 (\cite{dokkum94}) revealed extended light interpreted
as a weak thick disk with $h_{R}^{\rm thick} \simeq 2$~kpc.


Faint light high above the plane of the well-studied, late-type,
edge-on spiral NGC~5907 has further complicated the picture of
extra-planar light in small- or no-bulge spirals.
First detected at heights of 3 to 6~kpc
above the plane in deep $R$-band observations ({\cite{morrison94}),
this extended emission is intriguing because it is unlike any known
thick disk or stellar component, having instead a
morphology similar to that inferred for the
dark matter halo distribution of NGC~5907 (\cite{sackett94}). 
Other workers have confirmed the presence of 
the faint light in other bands (BVRIJK), and showed that the 
extended light is redder than the thin stellar disk.
If the faint light is due to a thick disk, it is unlike any other,
with a scale length that is at least twice that of its thin
disk (\cite{morrison99}).
The stellar population responsible for this faint light remains
highly controversial, ranging from normal or metal-rich populations
with steep IMFs (\cite{lequeux96}; \cite{rudy97}; \cite{jc98}), 
old, metal-rich accreted populations with normal IMF (\cite{lequeux98}), 
or exceedingly metal-poor or giant-poor populations with few resolvable
stars at the tip of the RGB (\cite{zepf00}).  
The controversy remains because the full spectral energy distribution is
apparently inconsistent with any single explanation 
(e.g. \cite{zepf00}; \cite{yost00}). 
\footnote{The discovery of a faint, long, very narrow arc
of light apparently associated
with NGC~5907 (\cite{shang98}) led Zheng et al. (1999) to suggest that
the extended light in the galaxy might be an
artifact due to confusion from the arc and foreground objects.
The arc clearly contributes
some light to some positions near the galaxy, but is too narrow and
asymmetric to be the cause of the symmetric extended light detected by
Morrison et al. (1994).
Zheng et al. (1999) report that their photometry suffers from
systematics at light levels fainter than $R = 27$\Msqarc.
(Due to a large pixel size,
the PSF was often undersampled, despite the seeing of
3.4 to 5.4\arcsec\ that was typical of their observations).
Since all detections of
faint extended light in NGC~5907 have been reported for
$R \geq 27$\Msqarc, and all optical photometry
(including that of \cite{zheng99}) agrees above this level,
the mystery of this faint halo light remains.}

The puzzling nature of the extended light in NGC~5907 has
motivated new studies to test a possible connection
between faint optical and IR light and dark matter in
this and other spirals
(\cite{gilmore98}; \cite{rauscher98}; \cite{uemizu98};
\cite{abe99}; \cite{beichman99}; \cite{yost00}; \cite{zepf00}).
The optical results are mixed, but infrared surface
brightness photometry
indicates that whatever produces the faint optical light
detected to date does not appear to emit strongly at IR
wavelengths far from the plane of the galactic disks.
Thus, if associated with known stellar populations, the
sources of the faint light are
unlikely to account for the dark mass of spiral galaxies.


\begin{table*}[t]
\begin{center}

\caption{Basic Properties of \gal}
\vspace*{0.5cm}

\begin{tabular}{c c c}
\hline
\noalign{\smallskip}
\multicolumn{1}{c}{Parameter}
&\multicolumn{1}{c}{Value}
&\multicolumn{1}{c}{Reference}\\
\hline\hline

$\alpha$,$\delta$ (J2000.0)	&21 12 10.8, $-$37 37 38	&\cite{karach99} \\
type				&Sc+6			&\cite{mathewson96} \\
redshift			&7680$\pm$ 10\kms	&\cite{mathewson96} \\
inclination			&88\degr			& this paper \\
PA				&120\fdg 4 $\pm$ 0\fdg 5	& this paper \\
major-axis D$^{\ast}$		&86\arcsec\			& this paper \\
m$_B$				& 16.67$\pm$0.09	&\cite{lauberts89}  \\
m$_V$				& 16.40$\pm$0.03		& this paper \\
m$_R$				& 15.92$\pm$0.04		& this paper \\
m$_I$				& 15.47$\pm$0.06	&\cite{mathewson96} \\
M$_R$				& $-$19.1$\pm$0.3			& this paper \\
M$_V$				& $-$18.7$\pm$0.3			& this paper \\
\noalign{\smallskip}
\hline
\end{tabular}
\end{center}

\footnotesize

\noindent{ }
\hspace*{3.4cm} $^{\ast}$major-axis diameter measured from the
R=27.0 \Msqarc\ contour.\\
\hspace*{3.8cm} Magnitudes are not corrected for extinction.

\label{tab:basic_props}
\end{table*}

In this paper, we report on the collection, reduction and
analysis of ultra-deep surface photometry of the isolated,
edge-on, low surface brightness, Sd galaxy \gal,
using some of the first science observations taken with the VLT.
The simple optics, good seeing, and
extremely well-sampled PSF of our observations ensured a low and
well-understood level of
scattered light and accurate identification of contaminating sources.  
Concurrent deep observations of unrelated blank fields with the VLT
were used to create dark sky flat fields at the appropriate wavelengths.
Considering all sources of uncertainty, including those from
light scattered through the wings of the PSF, we conclude
that the resulting surface photometry is reliable to a level of
$R = 28.5$\Msqarc\ and $V = 27.5$\Msqarc.
Analysis of these data reveals a faint component that we interpret as
a thick disk, to our knowledge the first thick disk discovered
in an LSB galaxy.

In Sect.~2 we describe the VLT observations and observing strategy.
In Sect.~3 the data reduction process, including the production of dark sky flats
and the procedures for masking, mosaicing, calibrating, and determining
the sky flux are outlined.
The procedure to extract profiles from the deep images
is given in Sect.~4, along with a brief description of the
error analysis, which is discussed in depth in the appendix.
The resulting $V$ and $R$ surface photometry of \gal\ are presented in
Sect.~5, along with a description of the
fitting procedure for the thin and thick disk parameters.
A thorough analysis of scattered light due to the tightly-constrained
PSF is discussed in Sect.~5, and ruled out as the
cause of the faint extended light we detect in \gal.
The thin and thick disks, including their inferred intrinsic
properties are described in Sect.~6.  We summarize and
conclude in Sect.~7.  Throughout this paper we assume a distance
of 102\,Mpc to \gal\ (based on a Hubble constant of  H$_\circ$=75\,km/s/Mpc),
which yields an image scale of 0.495\,kpc per arcsecond. 

\section{Observations}

\subsection{The Target Galaxy}

The target, \gal, is a nearby, edge-on galaxy,
selected on the basis of its right ascension and declination,
%
very high disk inclination,
absence of a prominent bulge, low extinction correction, and
optimal angular size.
The latter is important in order to adequately resolve
the disk scale height while maintaining sensitivity to faint
surface brightness in the halo.
Our deep $R$-band image obtained with
the VLT--UT1 test camera is shown in Fig.~\ref{fig:contour} and
the basic properties of the source are given in Table \ref{tab:basic_props}.

\begin{figure}[h]
\psfig{figure=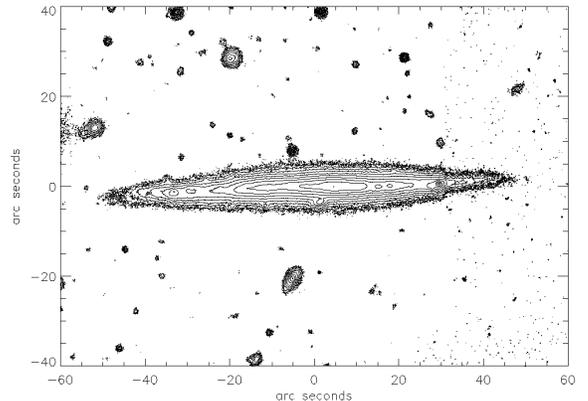,width=9.5cm}
\vspace*{-5cm}
\caption{Contour plot of \gal\ showing levels from 20.0 to 27.0
R~mag/sq~arcsec in 0.5~mag/sq~arcsec steps.  The image is a central
subsection of our total \gal\ mosaic.  The inability to trace
smooth contours at the lowest light levels and the
noisier background on the western end of the source is due
to fewer frames making up the mosaic on this side of \gal.}
\label{fig:contour}
\end{figure}

\subsection{Observing Strategy and Resulting Data} \label{sect:obs}

\begin{table*}[t]
\begin{center}

\caption{Summary of Observations}
\vspace*{0.5cm}
\begin{tabular}{l c c c c}
\hline
\noalign{\smallskip}
\multicolumn{1}{l}{Field}
&\multicolumn{1}{c}{Filter}
&\multicolumn{1}{c}{Dates}
&\multicolumn{1}{c}{Total Integration}
&\multicolumn{1}{c}{Median Seeing} \\
&& \multicolumn{1}{c}{August 1998}
&\multicolumn{1}{c}{(seconds)}
&\multicolumn{1}{c}{(arcsec)} \\
\hline\hline

\gal            &Bessel-$V$     &22                     &3300   &1\farcs 1   \\
                &Bessel-$R$     &18, 22, 25             &10320  &0\farcs 9   \\[4mm]

Flat-field frames: &&&& \\
&&&& \\

HDF--S$^a$      &Bessel-$V$     &18, 22, 23, 26, 27     &16200  &0\farcs 9   \\
                &Bessel-$R$     &18, 22, 23, 25, 26     &15300  &1\farcs 0   \\
EIS0046-2930    &Bessel-$V$     &17                     &2700   &0\farcs 8   \\
                &Bessel-$R$     &17                     &2700   &0\farcs 8   \\
EIS0046-2951    &Bessel-$V$     &22                     &2700   &0\farcs 9   \\


\noalign{\smallskip}
\hline
\end{tabular}
\end{center}
\noindent{ }
\label{tab:obslog}
\end{table*}
\normalsize

Bessel $V$ and $R$ observations of \gal\ were made on the nights of
18, 22 and 25 August 1998 as part of the ESO VLT--UT1 Science Verification
(SV) program.  A complete description of the VLT SV program
telescope and instrument set-up can be found in \cite{leib98} and
\cite{giacconi99}.  We give only a summary of the issues important
for our observations of \gal.

The VLT Test Camera, an engineering grade Tektronix
2048$^2$ CCD, was rebinned 2$\times$2 to improve its surface brightness
sensitivity, resulting in a binned scale of  0.091 arcsec~ pixel$^{-1}$
and a field of view of 93 arcsec on a side.
The camera was rotated approximately 60 degrees in order to position
the galaxy major axis along the x-axis of the detector.  For economy
of prose throughout the paper, we will refer to the
northeast and southwest sides of \gal\ as the
``northern'' and ``southern'' sides, respectively.

A challenge to our data reduction was the fact that the Test Camera
CCD is not a science-grade device.  As such, it displays more than
the customary number of cosmetic flaws, most noticeably, a large region
($\sim$120$\times$130 pixels) near the center of the chip with a
lower sensitivity than its surroundings.  Although this ``stain'' has a
strong colour dependence (it is more prominent in the blue), we found it
to be temporally stable and therefore easily corrected with our science
frame flat-fields (see Sect.~\ref{makeflat}).  Furthermore,
\gal\ was always positioned well away from this feature.

Fortuitously,
long total integrations were made of the HDF-south
and two EIS cluster candidate fields on the
nights of 17, 18, 22, 23, 25, and 26 August, in the same filters as
our observations.  Using these images to
create our deep sky flat-fields obviated the time-intensive strategy
of observing off-source fields for \gal.  Each of
the images used to make our superflats, as well as the observations of
\gal\ itself, were dithered on average by more than 10\arcsec\ in both
$\alpha$ and $\delta$.  This allowed for the removal of cosmic rays
from our galaxy field, and the removal of stars in the super
sky flat (see Sect.~\ref{makeflat}).

\section{Data Reduction} \label{reduce}

\subsection{Bias and Dark Current}

The basic image reduction was done using \mid.

The bias frames showed a fixed structure with an overall level that
varied up to 20 counts during the course of each night.  We
corrected for this by using the overscan region of the detector,
which mirrored the same variation.  For
each night, a median-filtered master bias was made from at least 20
individual bias images.  An average bias level was determined for each
image from its overscan region.  The associated master bias was then
scaled to each overscan mean and subtracted from each image, with the
0.5 count difference between the overscan and the bias average taken
into account.
No significant dark current was measured in the VLT test camera.

\subsection{Creating the Super Sky Flats} \label{makeflat}

The greatest potential source of error in our
final images is uncertainty in the flat-field.  
As many sky counts per pixel as possible are required to reduce the
statistical error in the flat-field which, to avoid large systematic
uncertainties, should be obtained using light with the
same spectral energy distribution as the
primary observation.  This was done by creating a super flat-field made 
from careful combinations of the deep EIS and the HDF-S fields that 
were interleaved temporally with our observations of \gal.  
The advantage of this method lies in the large total exposure of these
deep fields, which are devoid of bright stars and were well-dithered
between individual exposures.  
The HDF-S and EIS fields are located 26\fdg 3 and 53\fdg 8 away from \gal,
respectively.

Each candidate sky flat image was inspected visually; only
those free of defects and temporally close to our
observations of \gal\ were chosen.  Observations of the HDF-S made on
28, 29, and 31 August 1998 were not used in our flat-field due to
increasing sky levels from a waxing moon.
The remaining 26 $R$-band and 31 $V$-band flat frames contained
a total of
73560 and 39550 sky electrons per pixel, respectively.
Considering only Poisson statistics of sky electrons, the
flat-field formed from these frames should contribute a pixel-to-pixel 
error of 0.37\% ($R$-band) and 0.50\% ($V$-band).  Of course, variations
in the sky brightness across the image and remnant halos from inadequately
removed bright stars, create large-scale errors above that expected from
simple Poisson variations.  We empirically determine the size of this
dominant flat-field error below.

\begin{table*}[t]
\begin{center}

\caption{$RMS$ Flatness of Flat-fields}
\begin{tabular}{c c c c c c}
\hline
\noalign{\smallskip}
\multicolumn{1}{c}{Flat\/ Correcting Flat}
&\multicolumn{1}{c}{Filter}
&\multicolumn{1}{c}{Rebinned Size (\arcsec)}
&\multicolumn{1}{c}{Relevant Scale}
&\multicolumn{1}{c}{Measured $rms$}
&\multicolumn{1}{c}{$\frac{{\rm Pixel-to-Pixel}~rms}{\sqrt{N_{\rm pixel}}}$} \\
\hline\hline
$\mathrm{flatR1/flatR2}$  &$R$  &0\farcs 091 & 1 pixel & 0.57\% & -- \\
$\mathrm{flatV1/flatV2}$  &$V$  &0\farcs 091 & 1 pixel & 0.78\% & -- \\
\hline
$\mathrm{flatR1/flatR2}$  &$R$  &0\farcs 806 & 400\,pc ($\sim$h$_{\rm disk}$) & 0.11\% 
 & 0.064\% \\
$\mathrm{flatV1/flatV2}$  &$V$  &0\farcs 806 & 400\,pc ($\sim$h$_{\rm disk}$) & 0.14\% 
 & 0.088\% \\
\hline
$\mathrm{flatR1/flatR2}$  &$R$  &0\farcs 9   & 450\,pc (PSF FWHM in $R$) & 0.16\% 
 & 0.058\% \\
$\mathrm{flatV1/flatV2}$  &$V$  &1\farcs 1   & 550\,pc (PSF FWHM in $V$) & 0.12\% 
 & 0.065\% \\
\hline
$\mathrm{flatR1/flatR2}$  &$R$  &6\farcs 04 & 3\,kpc ($\sim$h$_{\rm halo}$) & 0.08\% 
 & 0.0086\% \\
$\mathrm{flatV1/flatV2}$  &$V$  &6\farcs 04 & 3\,kpc ($\sim$h$_{\rm halo}$) & 0.11\% 
 & 0.012\% \\
\noalign{\smallskip}
\hline
\label{tab:RMSflats}
\end{tabular}
\end{center}
\noindent{ }
\end{table*}

The super flat-field was created for each filter separately as follows.
Each individual flat-field sky frame was normalised to its
modal value as determined in the central 3/5 of the image.
The average value of pixel ($i,j$) was then determined from the stack of
sky frames for the filter, accepting a pixel ($i,j,k$) from the $k$th
frame in the computation of the average only if it passed two tests.
First, its deviation from the mean pixel value
in the stack at ($i,j$) must not exceed a given threshold measured
in units of the noise at that pixel position (a $\kappa$-$\sigma$ clip).
This criterion effectively removed cosmic-ray events and, since each
image was dithered by at least 10\arcsec\ (110 pixels) in both $\alpha$ and
$\delta$ between successive exposures,
the bright cores of stars and galaxies as well.  Second, a median-filtered
frame was created over a 3$\times$3 pixel window from the average frame
resulting from the first step.
A $\kappa$-$\sigma$ clip was again applied to each pixel ($i,j,k$)
based on the value of its local median.  The second test
was applied to remove any remnant faint extended wings of stars
and galaxies, which would otherwise contaminate the resulting
flat-field frame.  Only pixels satisfying both these ``filters''
entered the average for the flat-field frames.
A normalization level was calculated from the median value in
the central 3/5 of each flat-field frame, and each image
was then flattened and renormalized.

In order to test the quality of the flat-fields, and to compute an
empirical large scale flat-field error, we
repeated the above procedure using only one-half of the
available HDF-S and EIS images.
In this way, $flatR1$ was made from HDF-S and EIS images from nights
17, 18, 22, and 23 August, while $flatR2$ was made from HDF-S and EIS images
from nights 23 and 26 August.  The two subflats
$R1$ and $R2$ have approximately the same flux levels.
Two $V$-band subflats were created in the same way.  The flat-field frames
$flatR1$ and $flatV1$ were then flattened using $flatR2$ and $flatV2$,
respectively.  Each was then examined visually for any remnant
features, and then rebinned to a number of relevant scales and
the $rms$ variation across the frames measured.
The cosmetic flaws inherent in the
Test Camera CCD, particularly the ``stain'' mentioned in Sect.
\ref{sect:obs}, were removed effectively by our flat-field
procedure.  The results are summarized in Table~\ref{tab:RMSflats},
in which the measured $rms$ is compared to that expected from
photon statistics alone.  The empirical values
are used in our computation of flat-fielding errors.

\subsection{Mosaicing and Masking the Galaxy Frames} \label{make_mosaic_mask}

\begin{figure}[h]
\centerline{
\psfig{figure=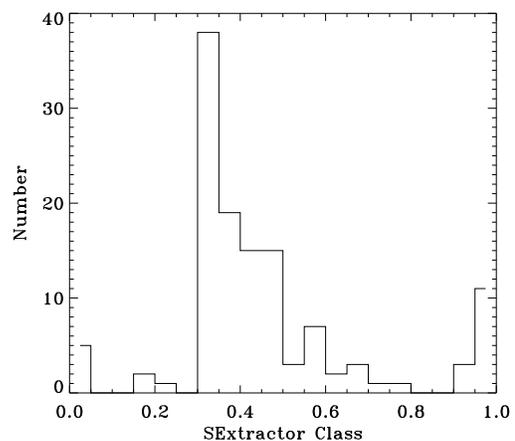,width=7.5cm}}
\caption{Histogram of the number of objects in our $R$-band image
detected by the \SeX\ program, as a function of the
object's classification.
The dividing line between a stellar and an extended detection
is approximately 0.8; the VLT field surrounding
\gal\ is clearly dominated by background galaxies.}
\label{fig:sex_detect}
\end{figure}

A region of sky 2\farcm8$\times$2\farcm2 ($R$) and 2\farcm2$\times$2\farcm0 ($V$)
around \gal\ was tiled with VLT test camera
exposures and then combined into a final mosaic.
Centroids of a number of stars and galaxies (usually 6 to
10) were measured in each individual image to
compute their positional offsets within the mosaic.
In order to remove cosmic ray events,
images were divided into groups of four closely overlapping frames.
Using the computed offsets, each group was combined into a temporary
median-filtered image.  Each input images was compared to
its group median and all pixels deviating by more than 3.5$\sigma$ were
replaced by the median value. Since cosmic ray events are often surrounded
by lower brightness halos or tails, a second iteration was done at each
position at which a cosmic ray was detected.  In this second pass, a lower
pixel correction criteria of 2.0$\sigma$ was applied.

The 14 ($R$-band) and 11 ($V$-band)
frames with the best seeing were then combined, using integer pixel
shifts, into $R$- and $V$-band mosaic frames.  Given the small pixel
size and large over-sampling, this did not limit the resolution of
our resulting image.  Since different regions of the
mosaic are constructed from different numbers of images, it is necessary
to renormalize.  To do this an identical set of frames was
created having the same sizes and offsets, but containing
only the modal value of the source-free sky background.
These were also combined into a mosaic
and used to renormalize the $R$- and $V$-band
mosaic frames.
A subsection of the $R$-band image resulting from this
procedure is shown in Fig.~\ref{fig:contour}.

\begin{figure}[t]
\vspace{0cm}
\hbox{\hspace{0cm}\psfig{figure=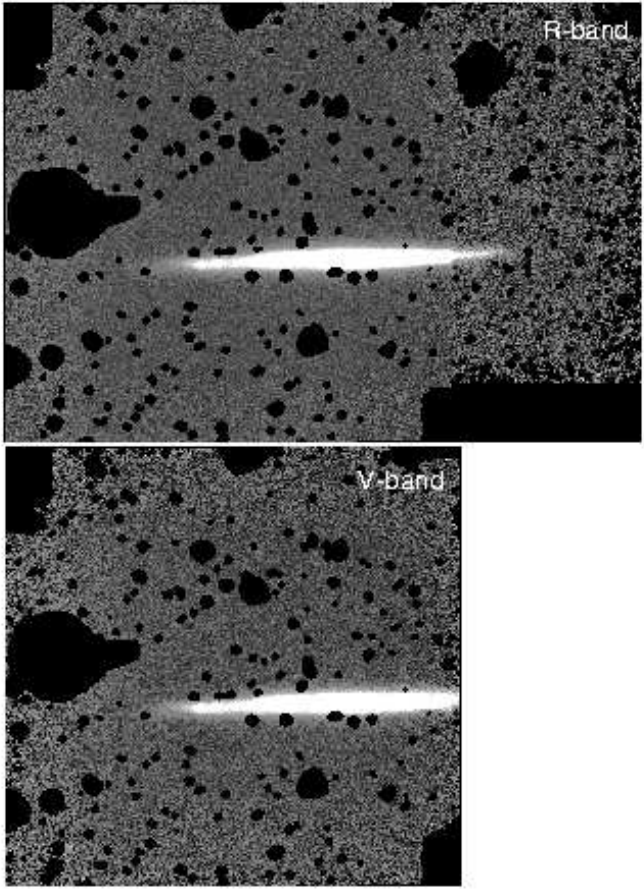,width=9.0cm}\hspace{0cm}}
\caption{Final $R$-band (top) and $V$-band masked images of \gal.
Objects detected with
\SeX\ in either band have been masked in both frames.
Levels $-3.5\sigma_{\mathrm sky}$ to
10$\sigma_{\mathrm sky}$ around the frame median are shown, where
$\sigma_{\mathrm sky}$ is the background rms/pixel of the frame.
}
\label{fig:objmaskRandV}
\end{figure}

In order to be able detect faint light associated
with \gal\ in our deep mosaic,
foreground stars and background galaxies must be masked out.
Since \gal\ was explicitly chosen for its paucity of
foreground stars, most of the objects contaminating its background
are galaxies (see Fig.~\ref{fig:sex_detect}), and simple
profile fitting cannot be used to model and subtract contaminants.

Instead, we used the \SeX\ detection algorithm (\cite{bertin}) to
find sources not associated with \gal.
A source was defined to consist of at least five connected pixels at a level of
1.5$\sigma$ above the local background, which was computed over
a 32$\times$32 pixel mesh.
The so-called \obj\ output of \SeX, essentially a frame of all
detected objects separated by null pixels, proved valuable in creating a
mask for objects beyond the outermost contours of \gal.
The initial output masks still retained a faint halo of emission around
brighter sources.  For this reason, the masks were
grown in size iteratively until a histogram of the unmasked background
pixels no longer changed shape, indicating that the local background
level had been reached.

A crucial step in the data reduction process is the determination
of an accurate value for the background sky level.  A
large central section of
both masked mosaics was extracted so that its area contained the
largest possible number of overlapping individual images
($\ge$11 for the $R$-band and $\ge$8 for the $V$-band).  In order to prevent
any emission from \gal\ contributing to the sky signal, the
galaxy was liberally masked out to 20\farcs 1 (10\,kpc) above
and below the central plane of its disk, and along its major axis to the
outermost edges of the images.  The mask sizes of
the brightest field stars were also liberally increased for this procedure.

\subsection{Determination of Sky Level}  \label{sect:skylevel}

\begin{figure}[h]
\centerline{
\psfig{figure=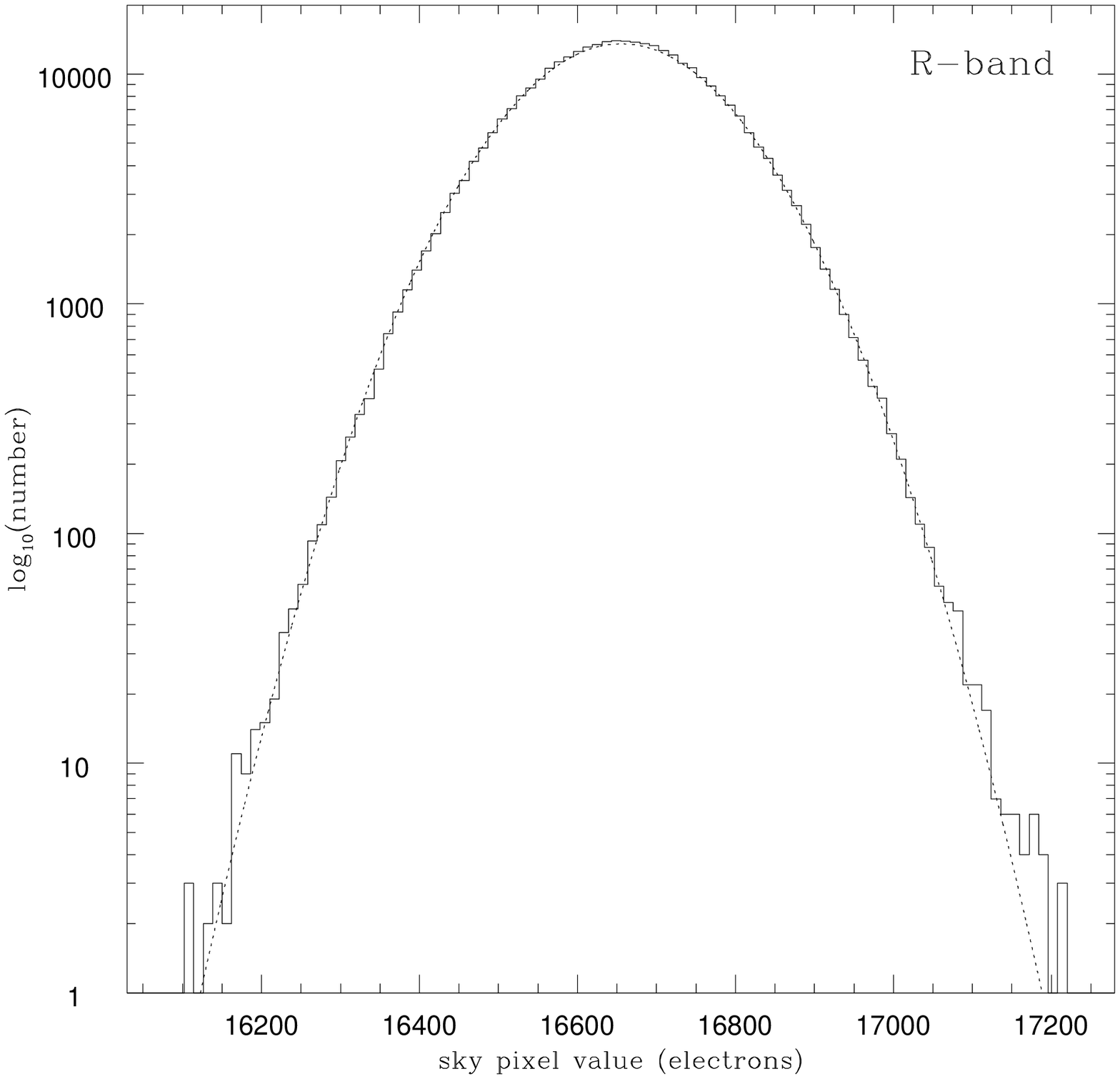,width=5.5cm,angle=0}}
\centerline{
\psfig{figure=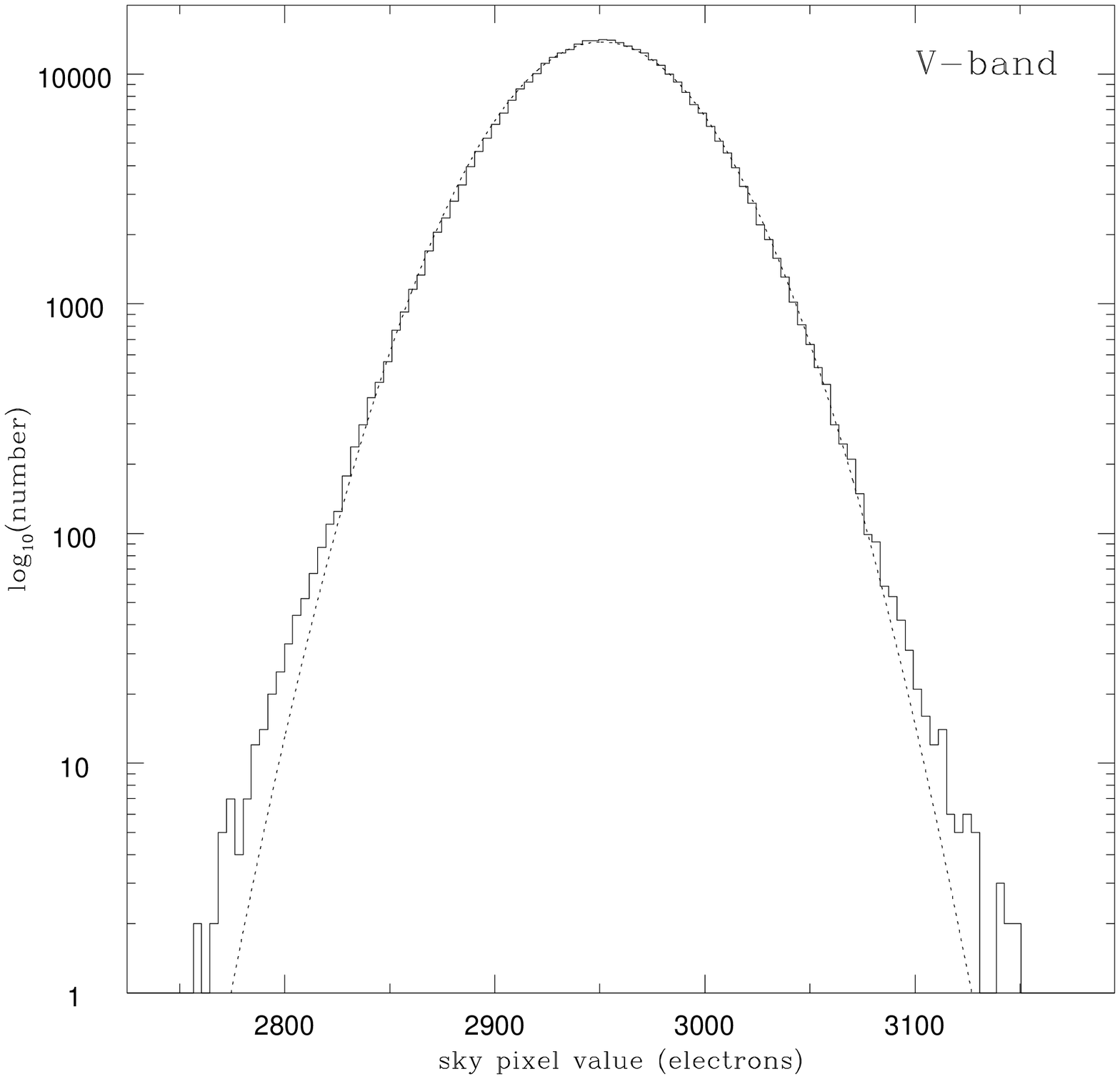,width=5.5cm,angle=0}}
\caption{Histogram of sky pixels from the completely masked $R$-band
(top panel) and $V$-band (bottom panel). A gaussian fit to the histograms
($exp(-(x-x_0)^2/2\sigma^2)$ with $x_0(R)$=16651.5 e$^{-1}$~pix$^{-1}$,
$\sigma(R)$=120.2 e$^{-1}$~pix$^{-1}$ and $x_0(V)$=2950.2 e$^{-1}$~pix$^{-1}$,
$\sigma(V)$=39.7 e$^{-1}$~pix$^{-1}$) is shown as a dotted line.
}
\label{fig:skyhist}
\end{figure}

\begin{figure*}[t]
\hbox{\hspace{-0.5cm}\psfig{figure=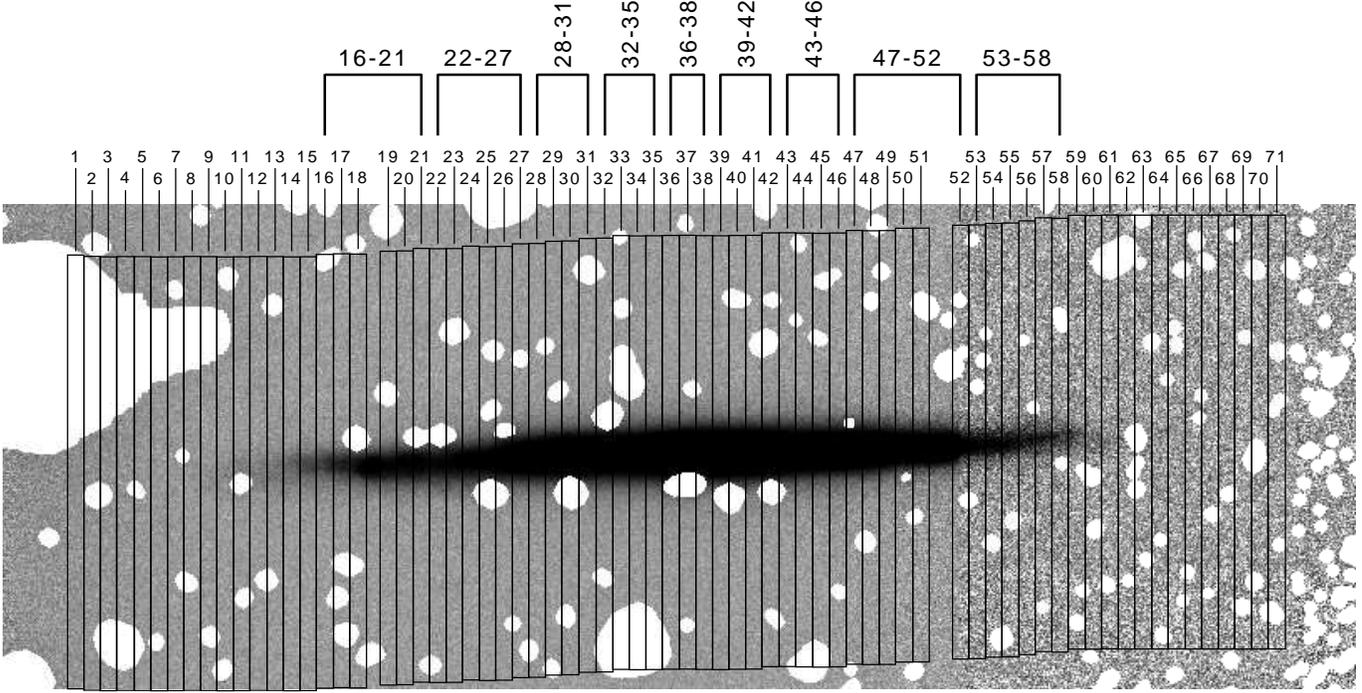,width=18cm}}
\caption{The positions of the profile extractions shown
on the mosaiced, masked, $R$-band image.
The $V$-band image was extracted at the identical
positions, but since it is smaller
(see Fig.~\ref{fig:objmaskRandV}), the $V$ profiles only reach number 52.
The vertical profiles averaged together to create Figs.~\ref{fig:profilesR}
and~\ref{fig:profilesV} are labelled at the top.}
\label{fig:vertcutR}
\end{figure*}

The distribution of sky values are shown in the histograms of
Fig.~\ref{fig:skyhist}, which were used to compute
the true background value of the unmasked pixels in each image, and the
associated error in its mean.  These sky values are
${\cal S}_{R} = {\rm 16651.5 \pm 0.4~e^{-}~pix^{-1}}$ and
${\cal S}_{V} = {\rm 2950.2 \pm 0.2~e^{-}~pix^{-1}}$
in the $R$ and $V$ bands, respectively.  Using the calibration
described in the next section, these values correspond to
m$_{\rm sky}$(R) = 20.98\Msqarc\ and m$_{\rm sky}$(V) = 21.60\Msqarc,
with a systematic uncertainty dominated by calibration errors of $\sim$5\%.
The systematic deviation from gaussian behaviour seen at extreme pixel values
in Fig.~\ref{fig:skyhist} is slight and very much smaller, in its
integrated effect on the average sky value, than the uncertainties
$\delta {\cal S}$ based on gaussian statistics reported above.

\subsection{Calibration to Standard System} \label{sect:photom}

Our photometric calibration was based on results supplied by the SV team
together with the distribution of our data.  A photometric solution
was available only
for the observations of \gal\ on 22 and 15 August, as these were the two
photometric nights.  Typically, four standard fields were observed several
times during each of these nights, with an average of
about 10 Landolt standard stars being used to compute the photometric
solutions.  The standards chosen spanned a significant range of colours
in order to adequately measure the colour term.

\section{Extracting Vertical Profiles} \label{sect:profiles}

Achieving acceptable signal-to-noise at surface
brightness levels 6 to 8 \Msqarc\ below sky requires
averaging over a large number of pixels.
We begin by extracting a number of vertical rectangular
regions, each of dimension
21 $\times$ 530 pixels (0.9$\times$ 24\,kpc), perpendicular
to the disk of \gal. These extracted areas are centered
on the major axis of the galaxy, avoid the most prominent HII regions,
and extend well beyond the visible disk.  The positions of the
extractions were identical for both the $R$ and $V$-band images;
71 regions were extracted from the $R$-band image and 52 from the
smaller $V$-band image.  Fig.~\ref{fig:vertcutR} shows these
areas atop on our masked mosaic $R$-band image.

\begin{table*}[t]
\begin{center}
\caption{Typical Errors for \gal\ Vertical Profiles}
\begin{tabular}{|l| c c| c c|}
\hline
\multicolumn{1}{|c|}{{\large Uncertainty}}
&\multicolumn{2}{|c|}{{\large R-band}}
&\multicolumn{2}{|c|}{{\large V-band}}\\
\multicolumn{1}{|c|}{[electrons]}
&\multicolumn{1}{|c}{Galaxy Center (\%)}
&\multicolumn{1}{c|}{$\pm$4\,kpc (\%)}
&\multicolumn{1}{|c}{Galaxy Center (\%)}
&\multicolumn{1}{c|}{$\pm$4\,kpc (\%)}\\[2mm]
\hline\hline

Averaged flux per pixel 			&30640 &16703 &5498 &2958 \\
Sky flux per pixel 				&16651.5 &16651.5 &2950.2 &2950.2\\
Net flux per pixel				&13988.5 &51.5 &2547.8 &7.8
\\[2mm]

Read Noise ($\sigma_{\rm RN}$) 			&0.04 (0.0003) &0.04 (0.08)
						&0.05 (0.002) &0.05 (0.6) \\

Flat-Fielding ($\sigma_{\rm FF}$) 		&7.4 (0.05) &4.0 (7.8)
						&1.8 (0.07)   &1.0 (12.5) \\
Photon Noise ($\sigma_{\rm PN}$)  		&1.0 (0.007)  &0.8 (1.5)
					 	&0.5 (0.02)  &0.4 (5.4) \\
Mosaicing Error ($\sigma_{\rm M}$) 		&1.7 (0.01)  &0.9 (1.8)
					 	&2.6 (0.1)  &1.4 (17.9) \\[2mm]
Surface Brightness Fluctuations ($\sigma_{\rm L}$) &0.2 (0.001) &0.01 (0.02)
					 	&0.06 (0.002) &0.003 (0.04)\\[2mm]
Total Statistical Error ($\sigma_{\rm STAT}$) 	&7.6 (0.05) &4.2 (8.1)
					 	&3.2 (0.1)   &1.7 (22.3) \\[2mm]
Sky Subtraction ($\sigma_{\rm SS}$)$^{\ast}$  	&0.4 (0.003) &0.4 (0.8)
					 	&0.2 (0.008) &0.2 (2.6) \\
Total ${\rm m \pm \Delta m}$ [\Msqarc]  	&21.17 $\pm$ 0.04
					 	&27.3 $^{+0.13}_{-0.12}$
					 	&21.76 $\pm$ 0.03
					 	&28.0 $^{+0.30}_{-0.25}$ \\[2mm]
\hline
\end{tabular}
\label{tab:errtab}
\end{center}
 \footnotesize
\noindent{ }
The numbers in parentheses are the errors as a percentage of the sky-subtracted flux.\\
$^\ast$ Note:  the sky subtraction error is a systematic error
and, therefore, does not enter into $\sigma_{\rm STAT}$.
\end{table*}

From these initial extractions four levels of averaging were performed
in order to increase the signal-to-noise:
\begin{itemize}
\item[1.] The sum of the flux across the 21 pixel wide $x$-direction was
determined for each extraction, and normalized by the number of non-masked
pixels contained in each row.
\item[2.] Due to the extremely symmetrical cross-section of \gal\
(see Fig.~\ref{fig:NSsym}) we were able to average the profiles above
and below the disk.
\item[3.] In the vertical direction ($z$-axis), each profile was averaged
over the size of the seeing FWHM ($\sim$10 pixels) for all points
above the plane of the galaxy.
\item[4.] Finally, a number of profiles were averaged together.  An average
of the three innermost profiles (extractions 36 to 38)
was made to create one central profile, groups of four profiles
(28--31 and 32--35 to the east and 39--42 and 43--46 to the west) were
averaged on each side of the center, groups of six
(16--21 and 22--27 to the east and 47--52 and 53--58 to the west) were
averaged on the outermost ends of the galaxy.  The extractions averaged
together are shown at the top of Fig.~\ref{fig:vertcutR}.
In the case of the $V$-band image, the averaging process ends with extraction 
number 52.
\end{itemize}

The resulting masks made from the $R$ and
$V$ images separately, were then multiplied together to create a
master mask frame that was applied to each mosaic.
This procedure masked 10.3\% and 11.0\% of the total image areas
in the final $R$ and $V$ mosaics respectively.
The masked images are shown in Fig.~\ref{fig:objmaskRandV}.

\begin{figure}[h]
\resizebox{\hsize}{!}{\includegraphics{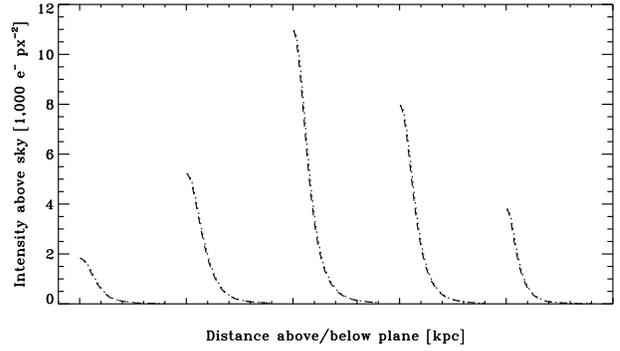}}
\caption{The symmetry of the vertical surface brightness profiles of \gal.
The north (dashed lines) and south (dotted lines) profiles extracted
at various positions along the disk.  The object-masked images have been used.
}
\label{fig:NSsym}
\end{figure}

For each of the vertical extractions covering the
visible disk of \gal\ (profiles 16 to 58 in $R$ and 16 to 52 in $V$), a
least-squares fit to the thin disk component was made.
A simultaneous two-component (thin and thick disk) fit was made to each
extraction, using an exponential parametrization given by
\begin{equation}
f(z) = f_\circ^{\rm thin} exp(-|z|/h_z^{\rm thin}) + 
f_\circ^{\rm thick} exp(-|z|/h_z^{\rm thick})
\end{equation}
for both components,
where $f_\circ$ is the surface flux at the position the extraction
crosses the major axis of the galaxy,
$z$ is the projected distance from the major axis,
and $h_z$ is the exponential scale height.  
The fitted parameters are $f_\circ^{\rm thin}$, $h_z^{\rm thin}$, 
$f_\circ^{\rm thick}$, and $h_z^{\rm thick}$.
Regions strongly affected by
dust or clumpy HII regions were excluded from the fit.
For comparison, a single-component (thin disk) fit was also performed for
those data that lie between 1 and 3\Msqarc\ below the central galaxy
surface brightness.  Results
are presented in Sect.~\ref{sect:results}.

\subsection{Error Analysis} \label{sect:errors}

We present here a brief summary of the sources of photometric
uncertainty and their magnitudes; the reader is referred to the Appendix
for a more detailed discussion.

For illustration, Table~\ref{tab:errtab} shows the average flux levels
and uncertainties at two positions along the central vertical profile
(the average of extractions 36 to 38)
of \gal, one at the
galaxy center and another at the much fainter light levels
4\,kpc above the galaxy disk.
For each of the two flux extremes in each of $R$ and $V$-bands,
we give the uncertainties associated with the average flux per
pixel (averaged over the unmasked area of 3$\times$21$\times$10 pixels) in
units of electrons and as a percentage of the {\em sky-subtracted} flux
(given in parentheses).

\begin{figure*}[t]
\vspace{0cm}
\hbox{\hspace{0cm}\psfig{figure=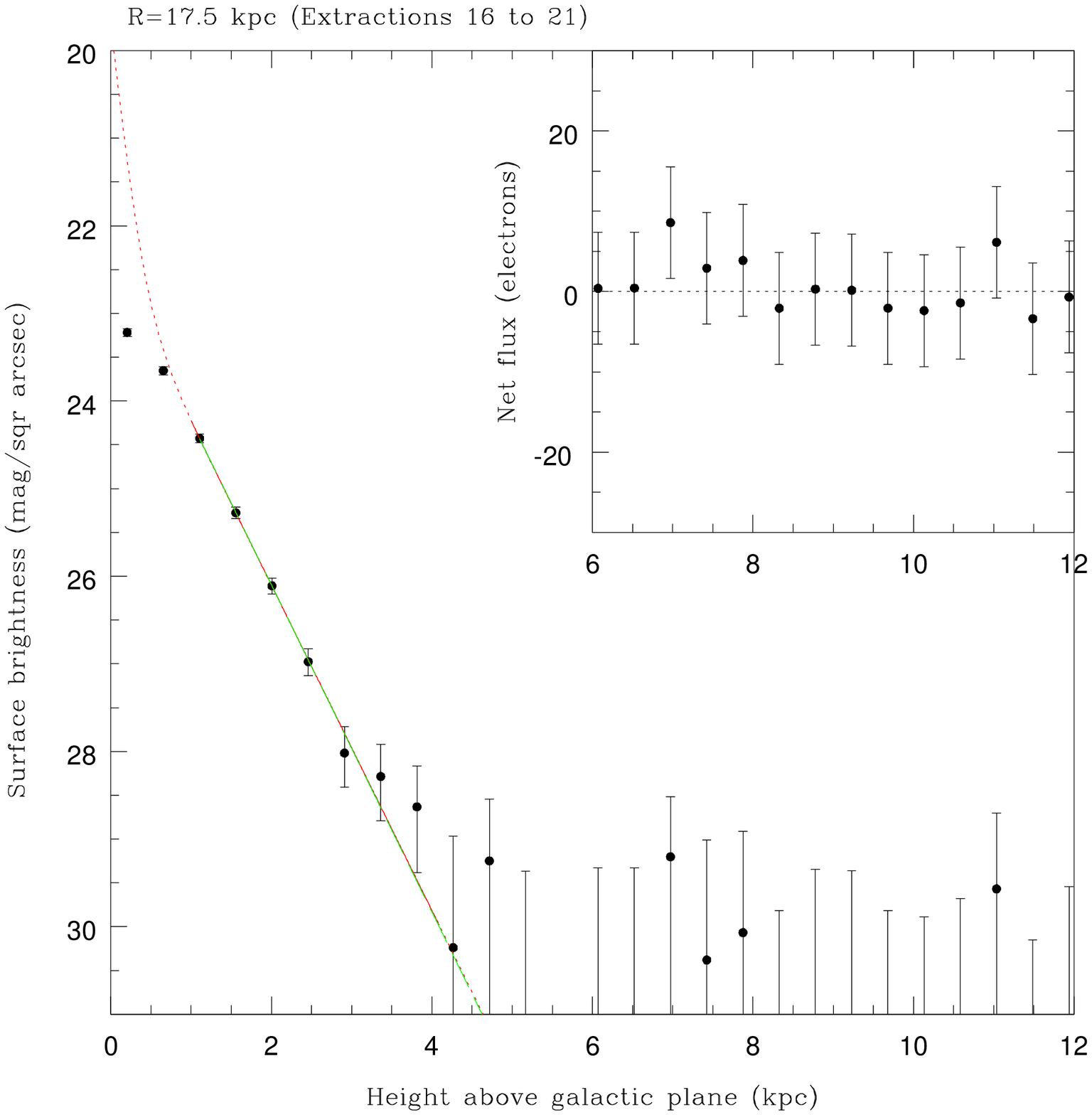,width=6.0cm}\hspace{0cm}
\psfig{figure=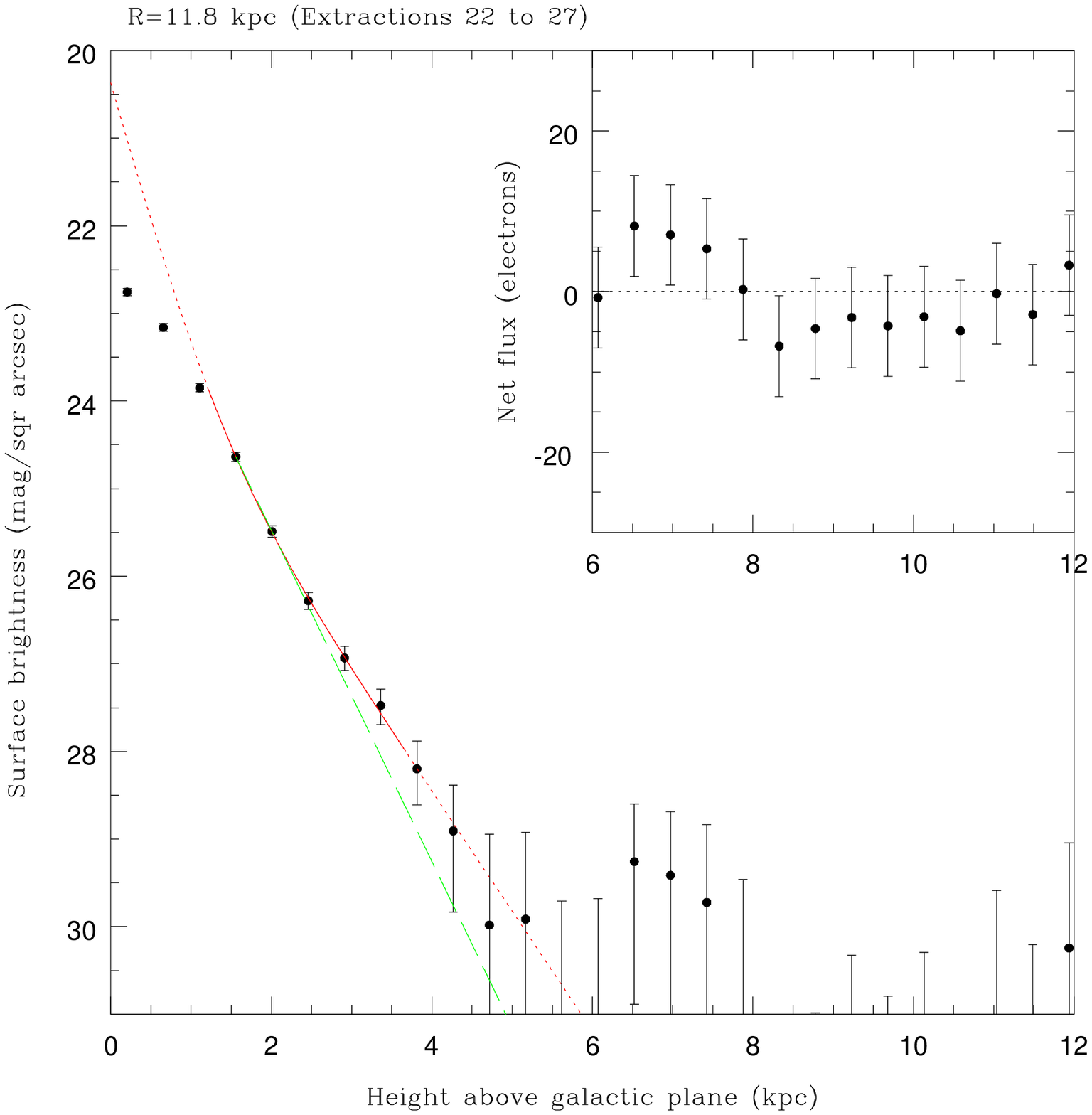,width=6.0cm}
\psfig{figure=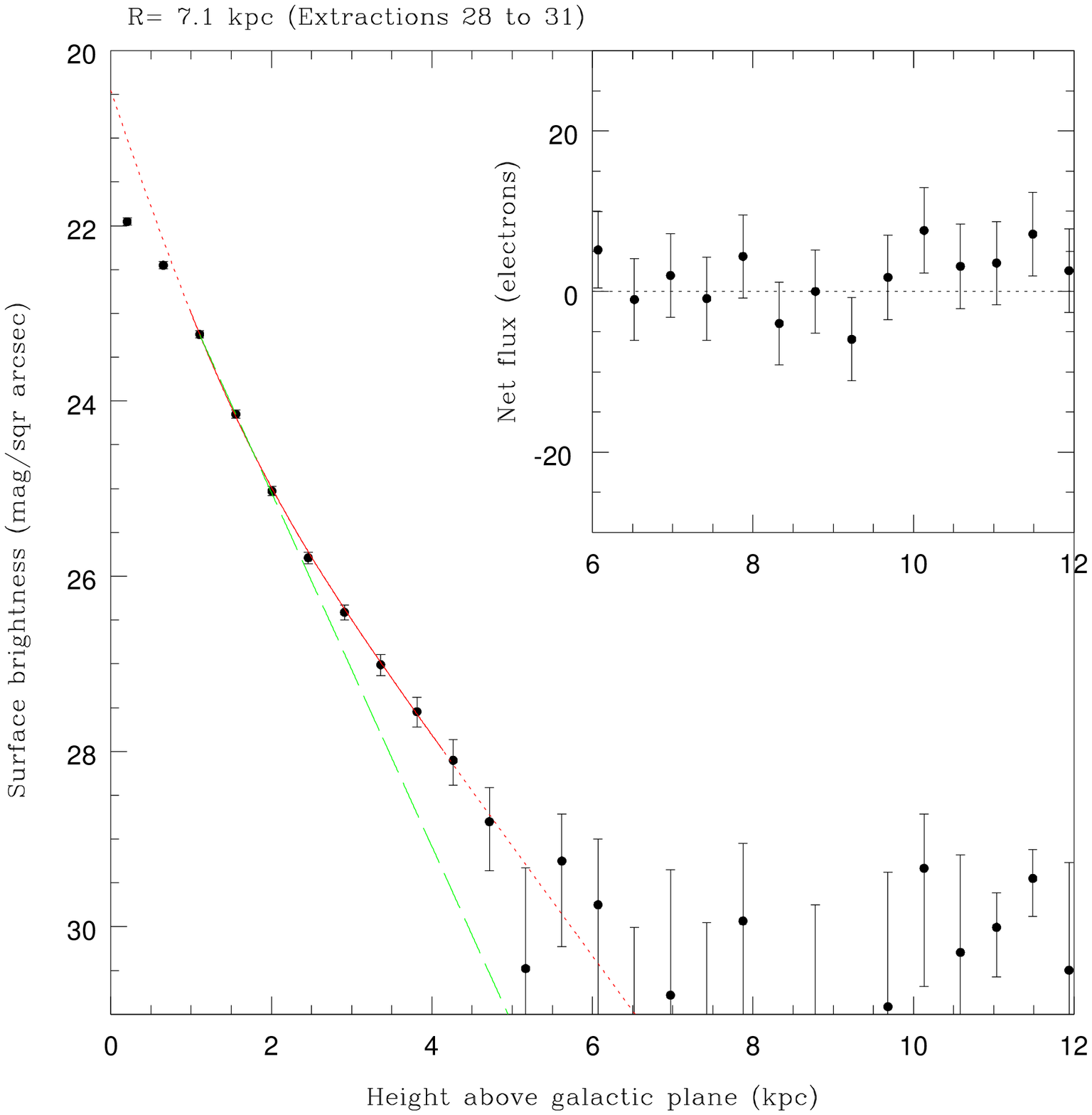,width=6.0cm}}
\hbox{\hspace{0cm}\psfig{figure=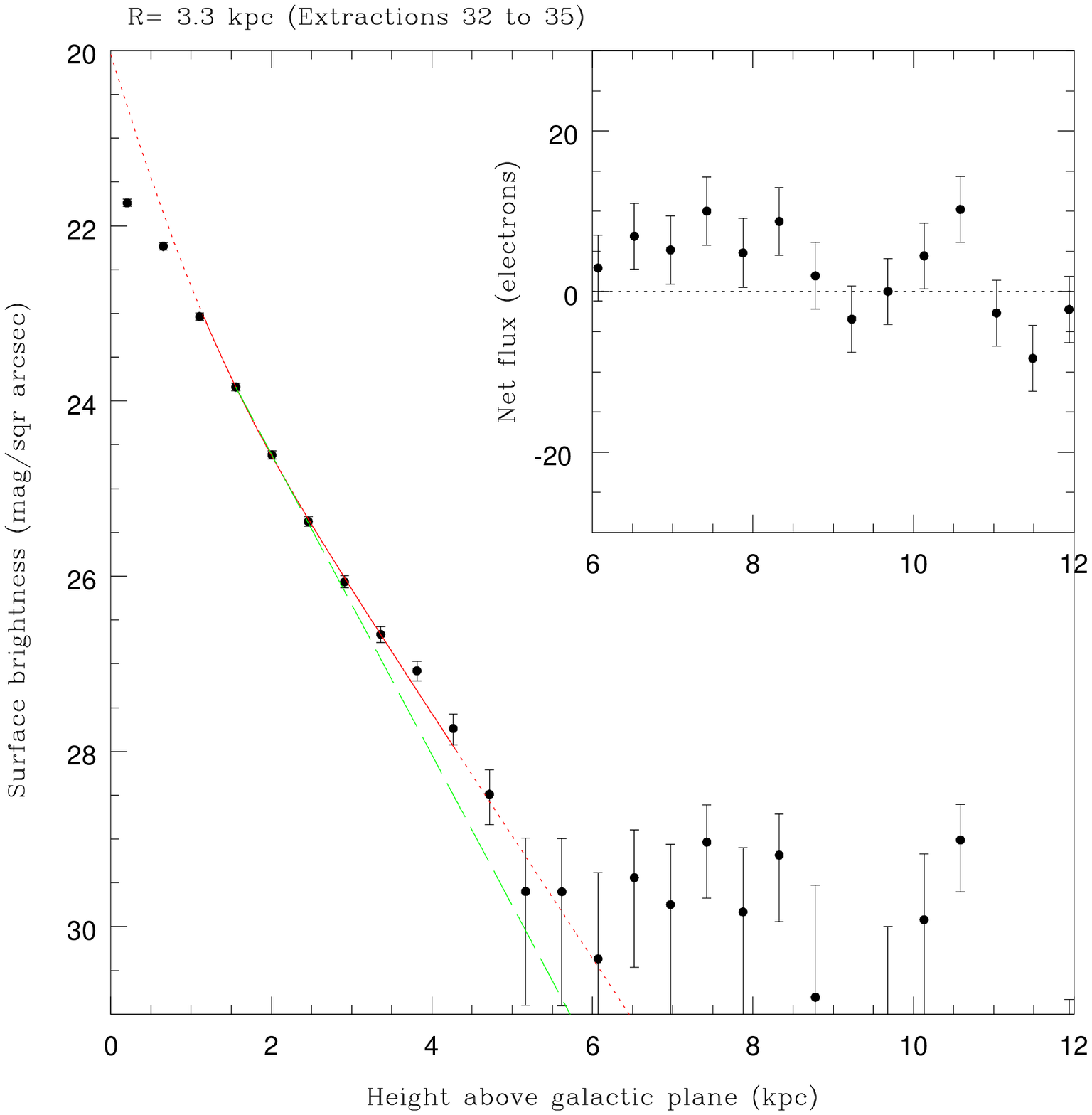,width=6.0cm}\hspace{0cm}
\psfig{figure=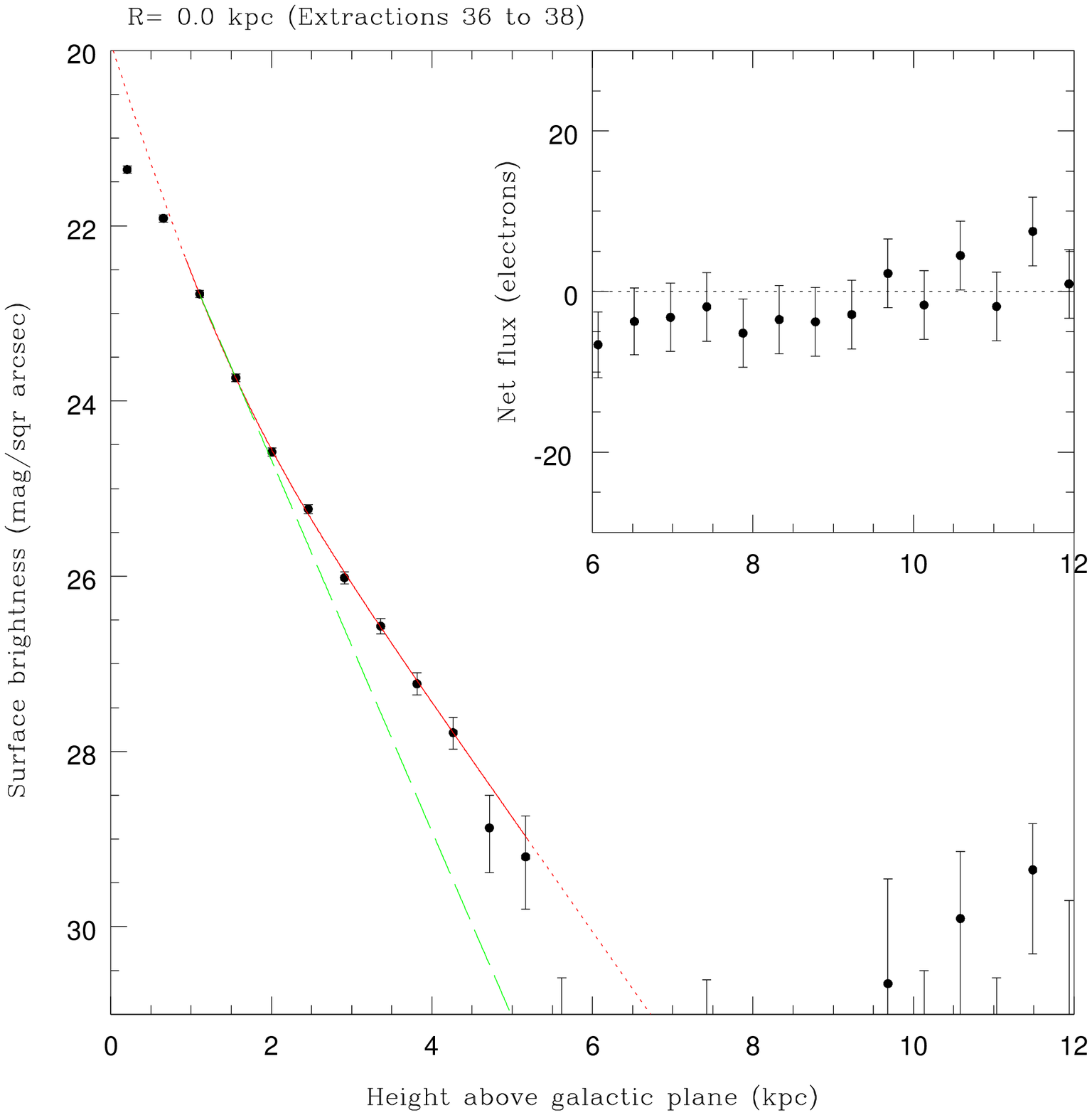,width=6.0cm}
\psfig{figure=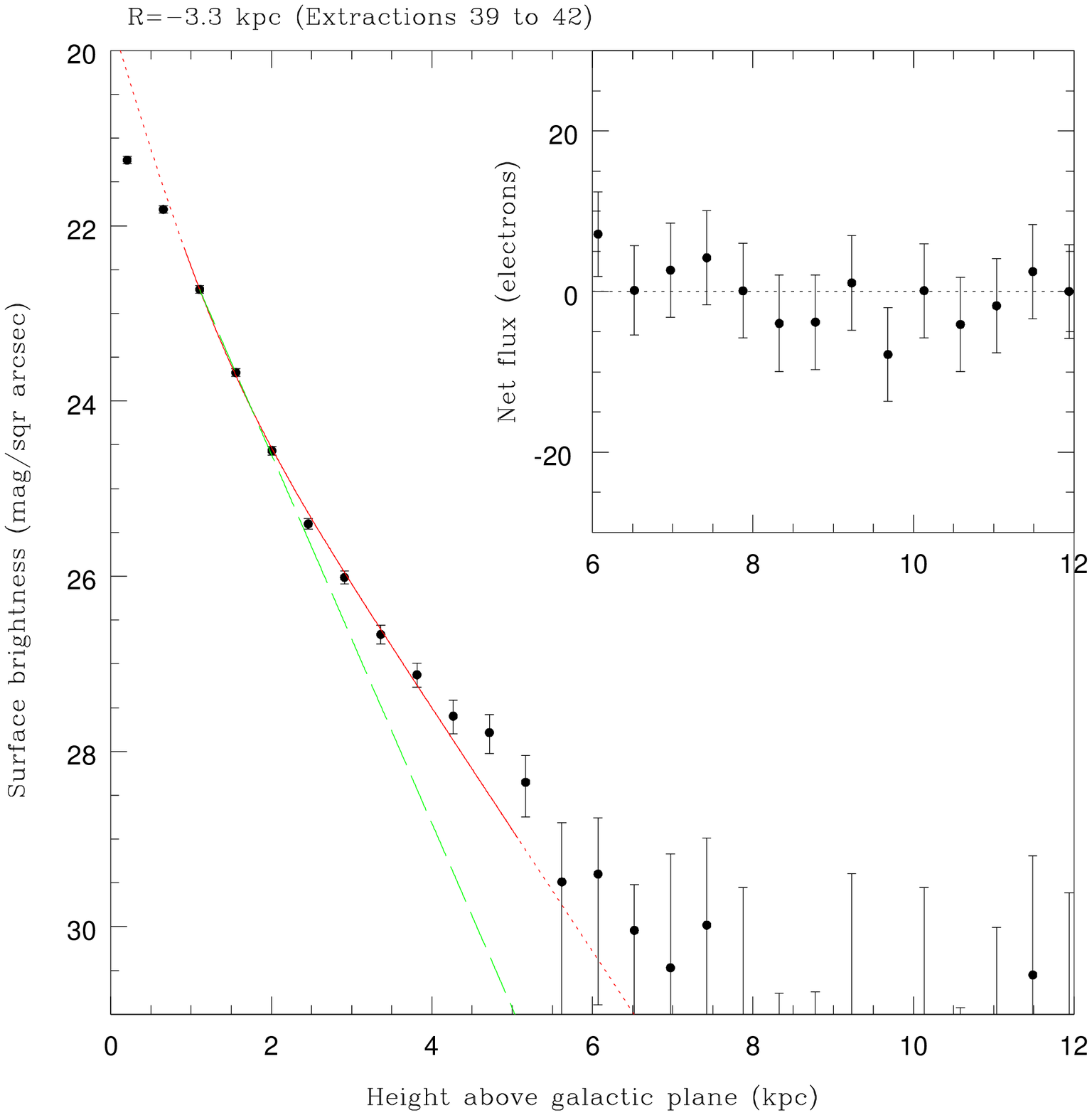,width=6.0cm}}
\hbox{\hspace{0cm}\psfig{figure=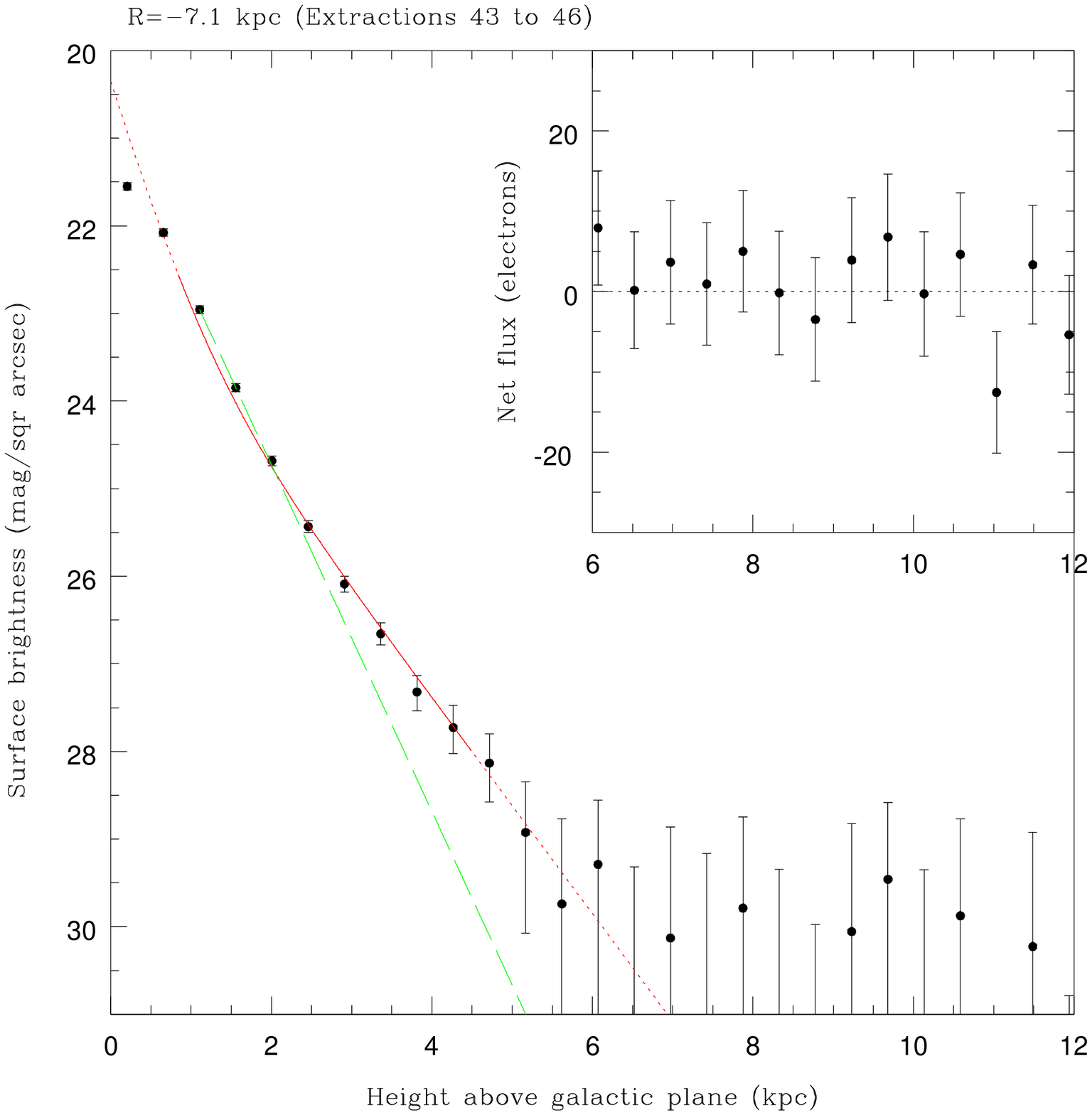,width=6.0cm}\hspace{0cm}
\psfig{figure=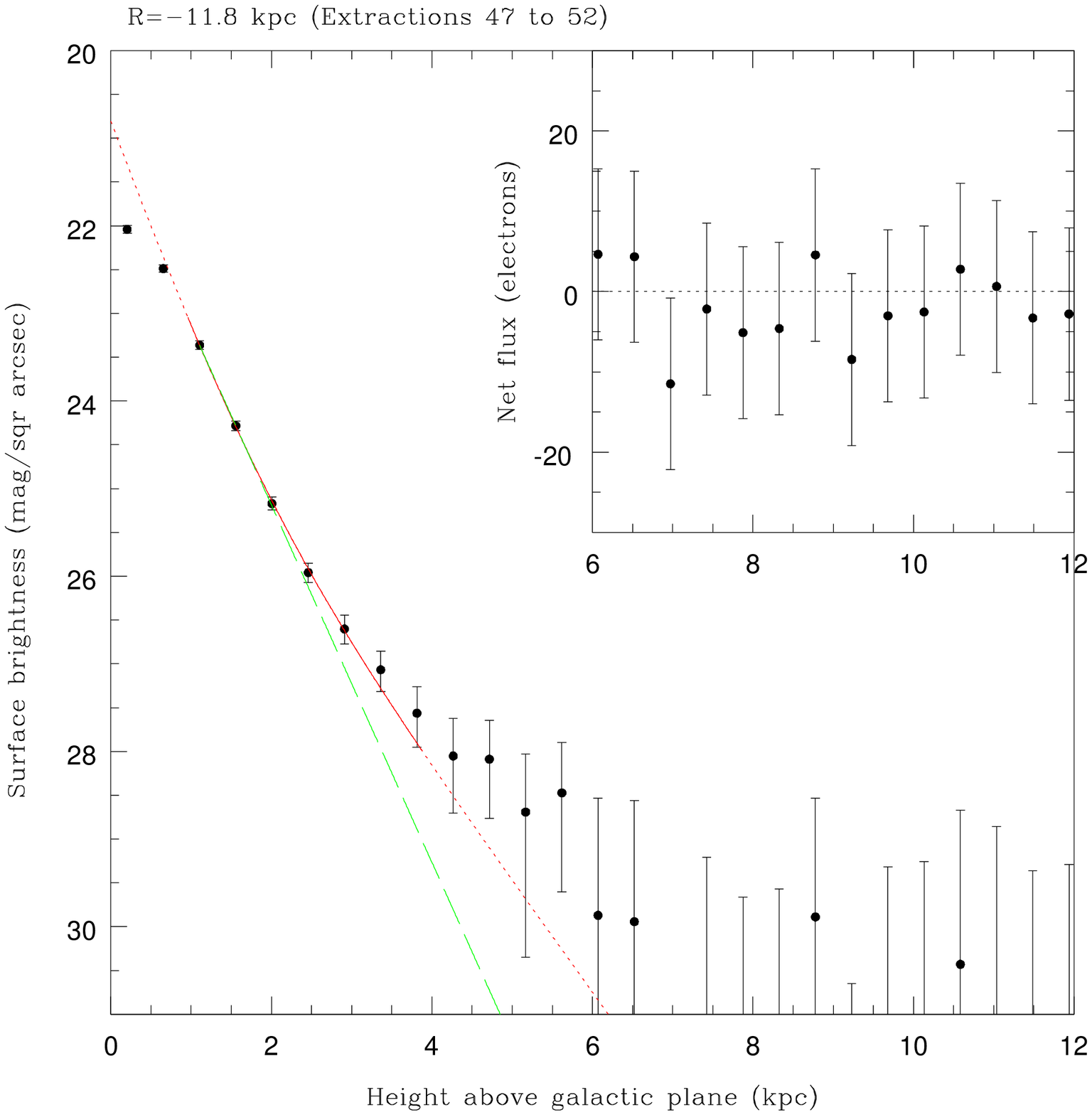,width=6.0cm}
\psfig{figure=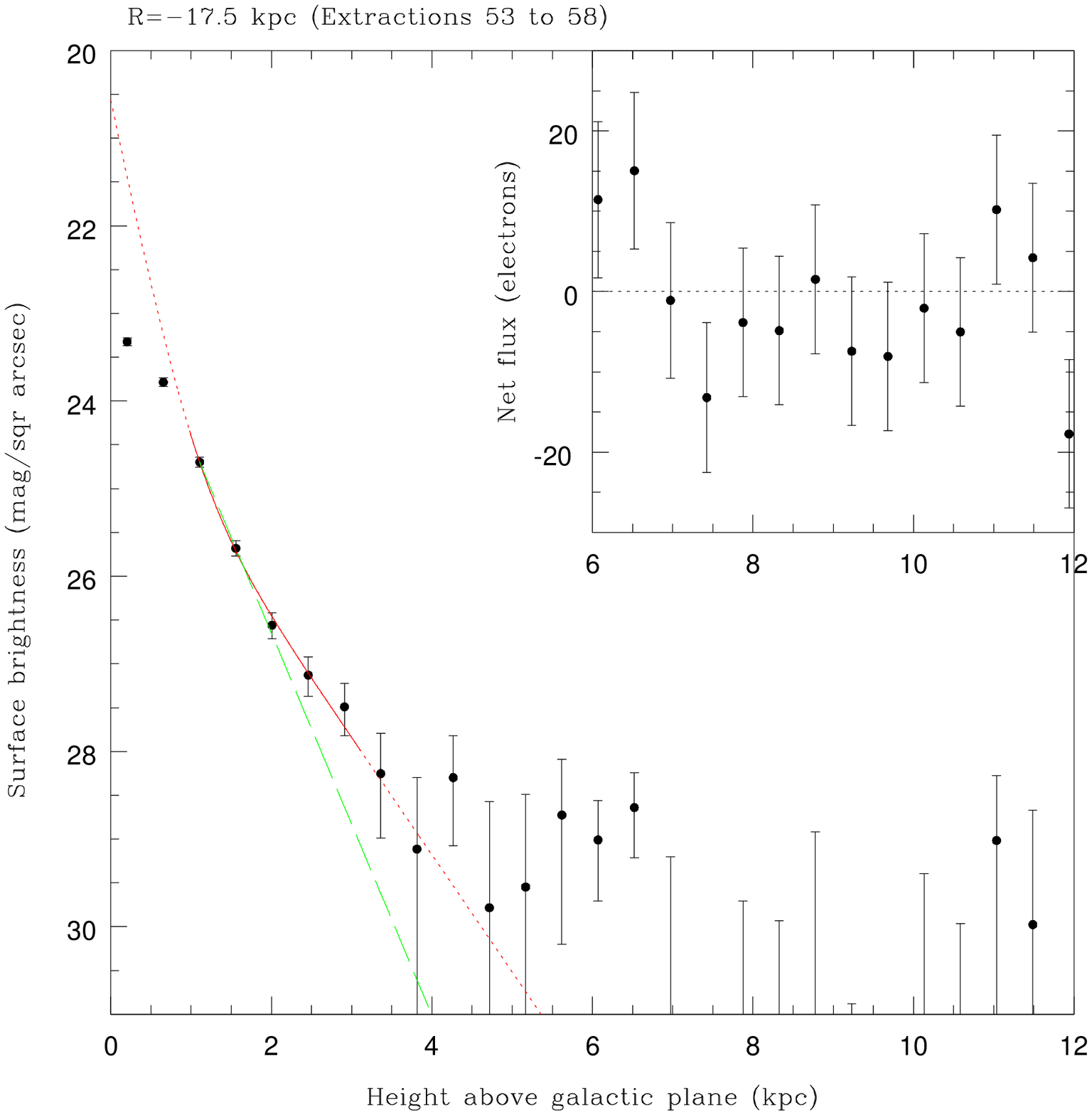,width=6.0cm}}
\vspace{0cm}
\caption{The $R$-band averaged profiles through the disk of \gal,
perpendicular to the major axis.
The average position from the galaxy
center is given at the top of each panel (east of center is
indicated by $R>$0; west of center by $R<$0).
The insets show, on a linear scale, the background-subtracted flux
levels of each profile at distances more than 6\,kpc from the galaxy disk.
The simultaneous thin and thick disk fit and the range of data
used for the fit is shown by a solid line;
the dotted line is the extrapolation of the fit.
The dashed line is intended as a guide, and indicates a single-component fit to
data dominated by the thin disk.  This fit was restricted to data between
1 and 3 mag/sqr arcsec fainter than the peak flux.
}
\label{fig:profilesR}
\end{figure*}

\begin{figure*}[t]
\vspace{0cm}
\hbox{\hspace{0cm}\psfig{figure=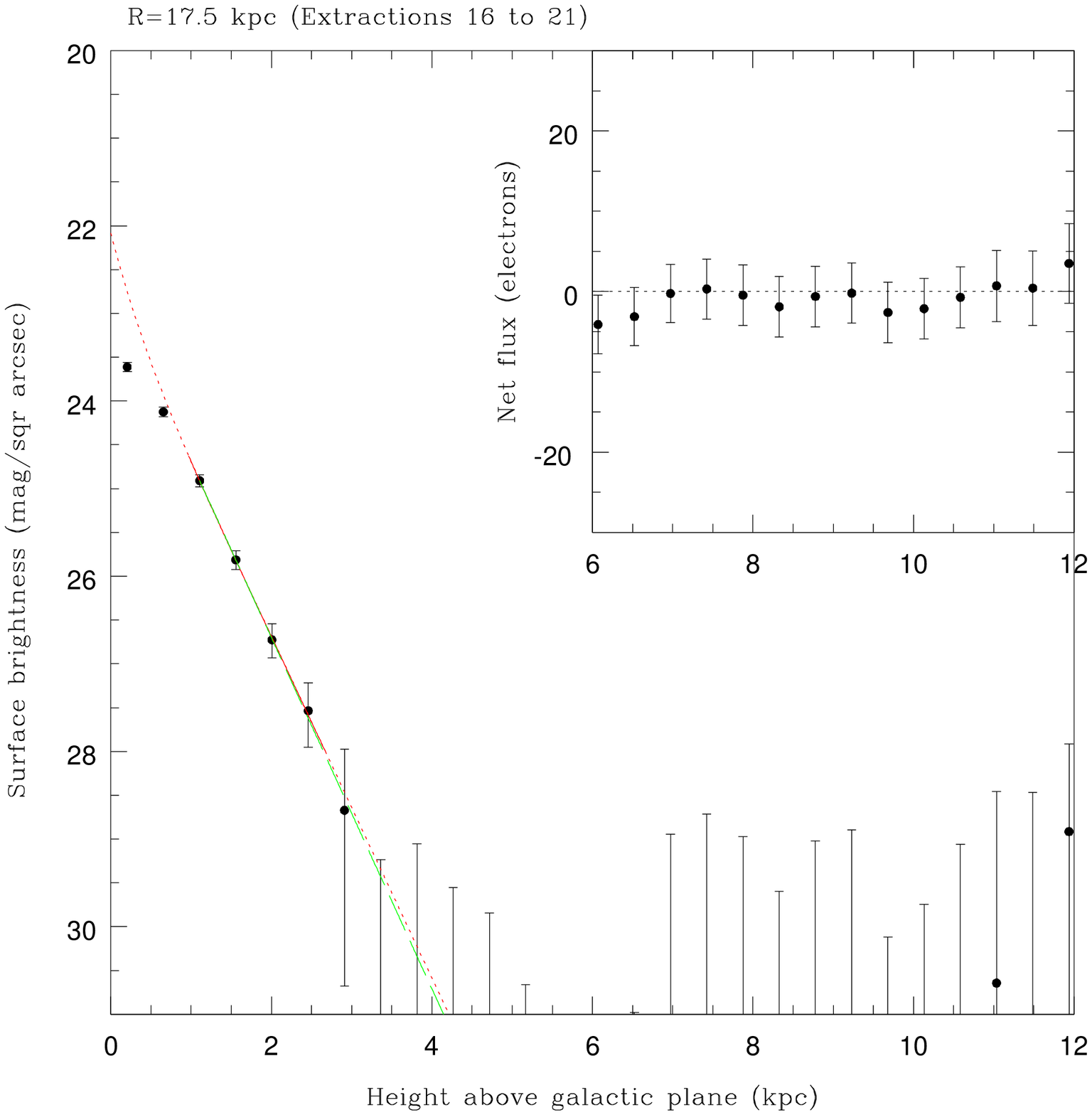,width=6.0cm}\hspace{0cm}
\psfig{figure=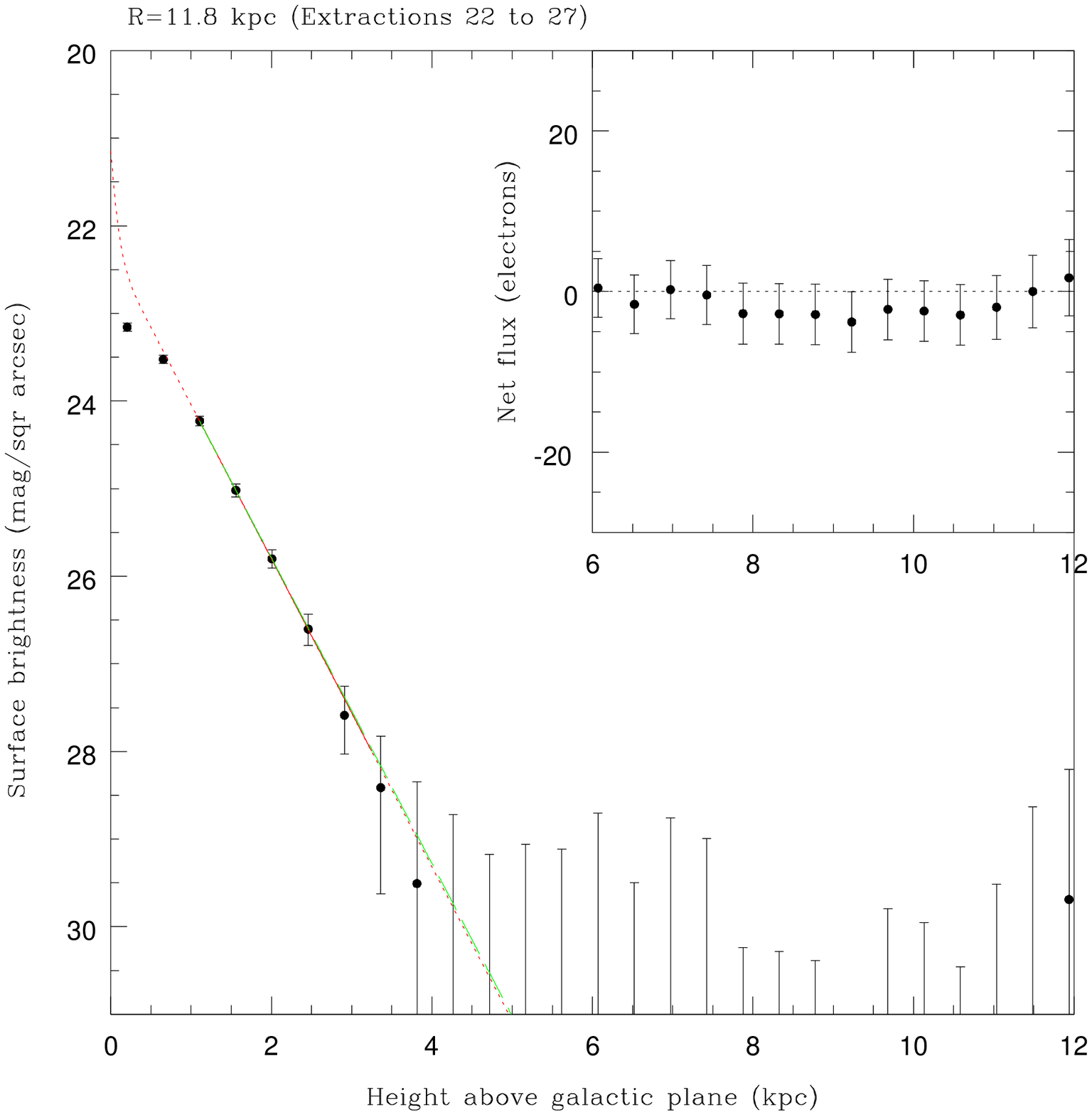,width=6.0cm}
\psfig{figure=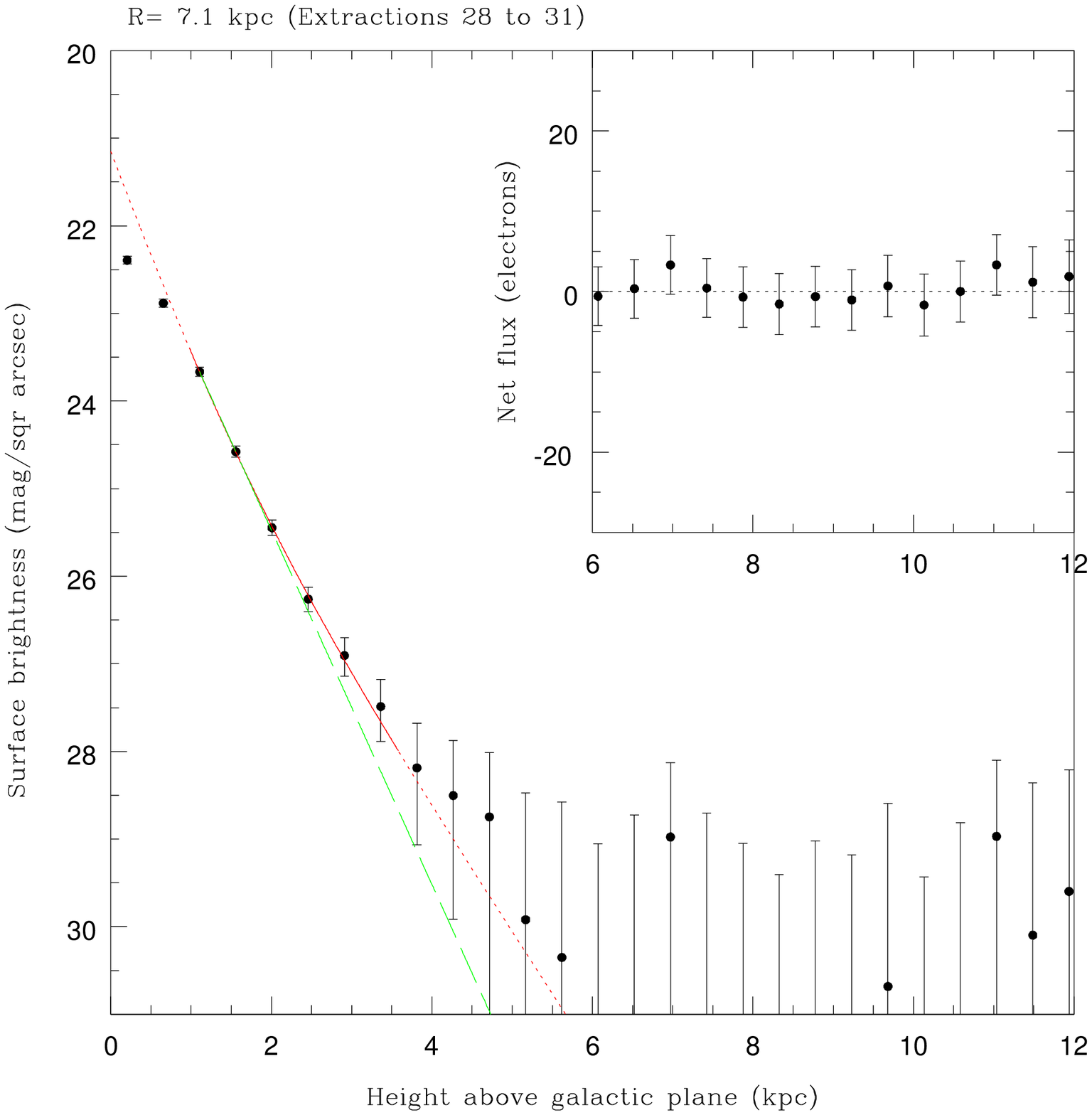,width=6.0cm}}
\hbox{\hspace{0cm}\psfig{figure=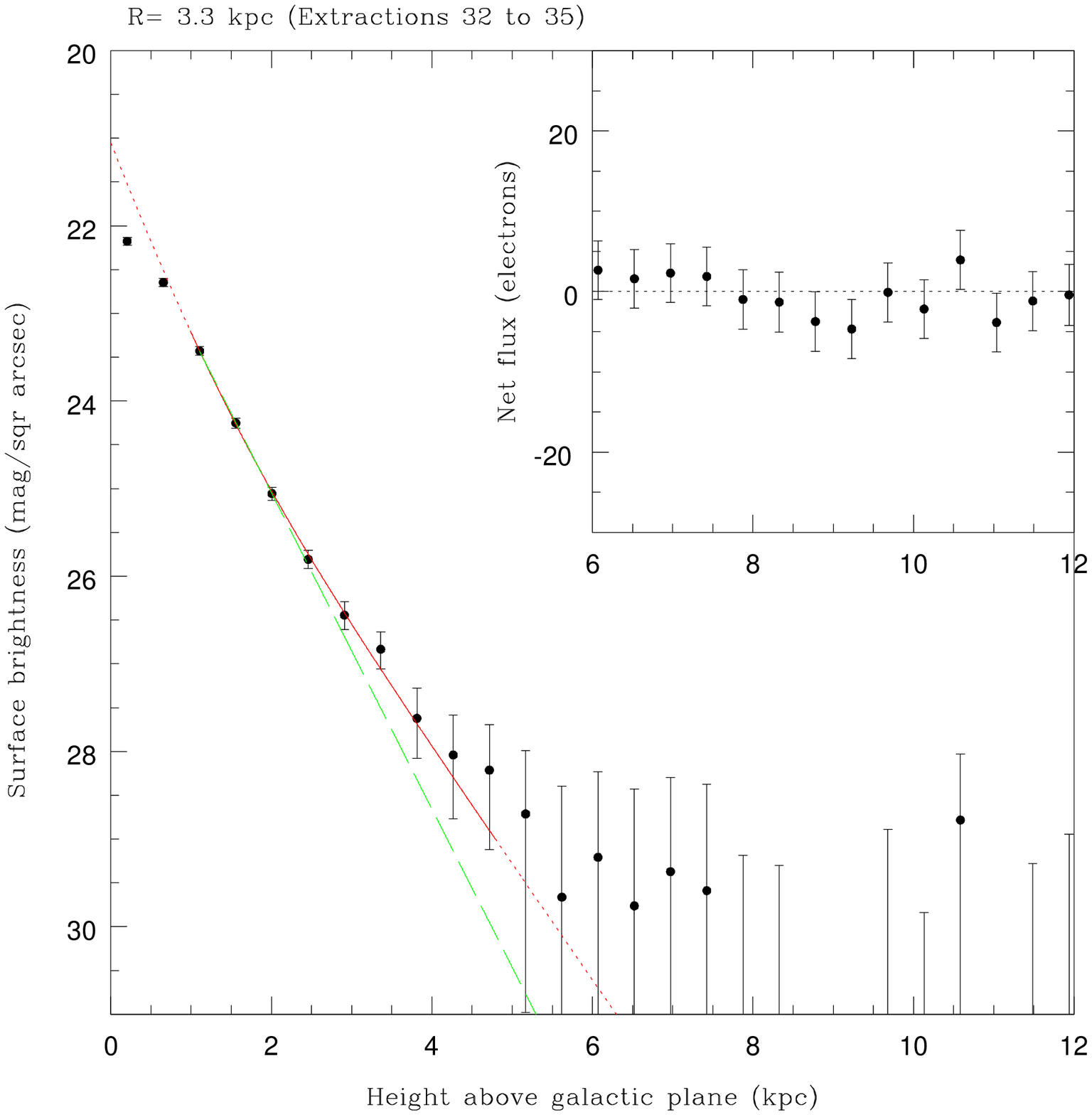,width=6.0cm}\hspace{0cm}
\psfig{figure=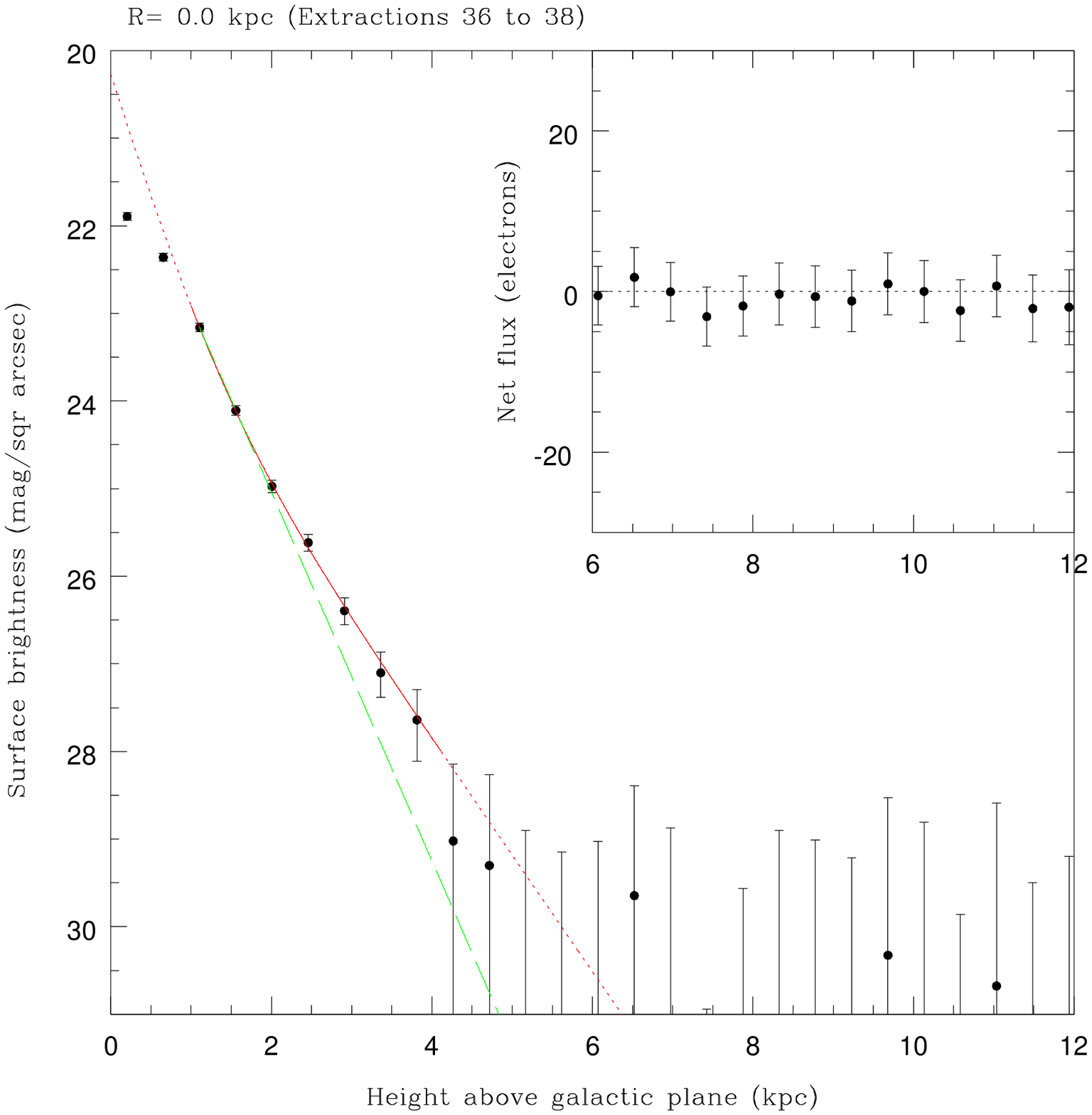,width=6.0cm}
\psfig{figure=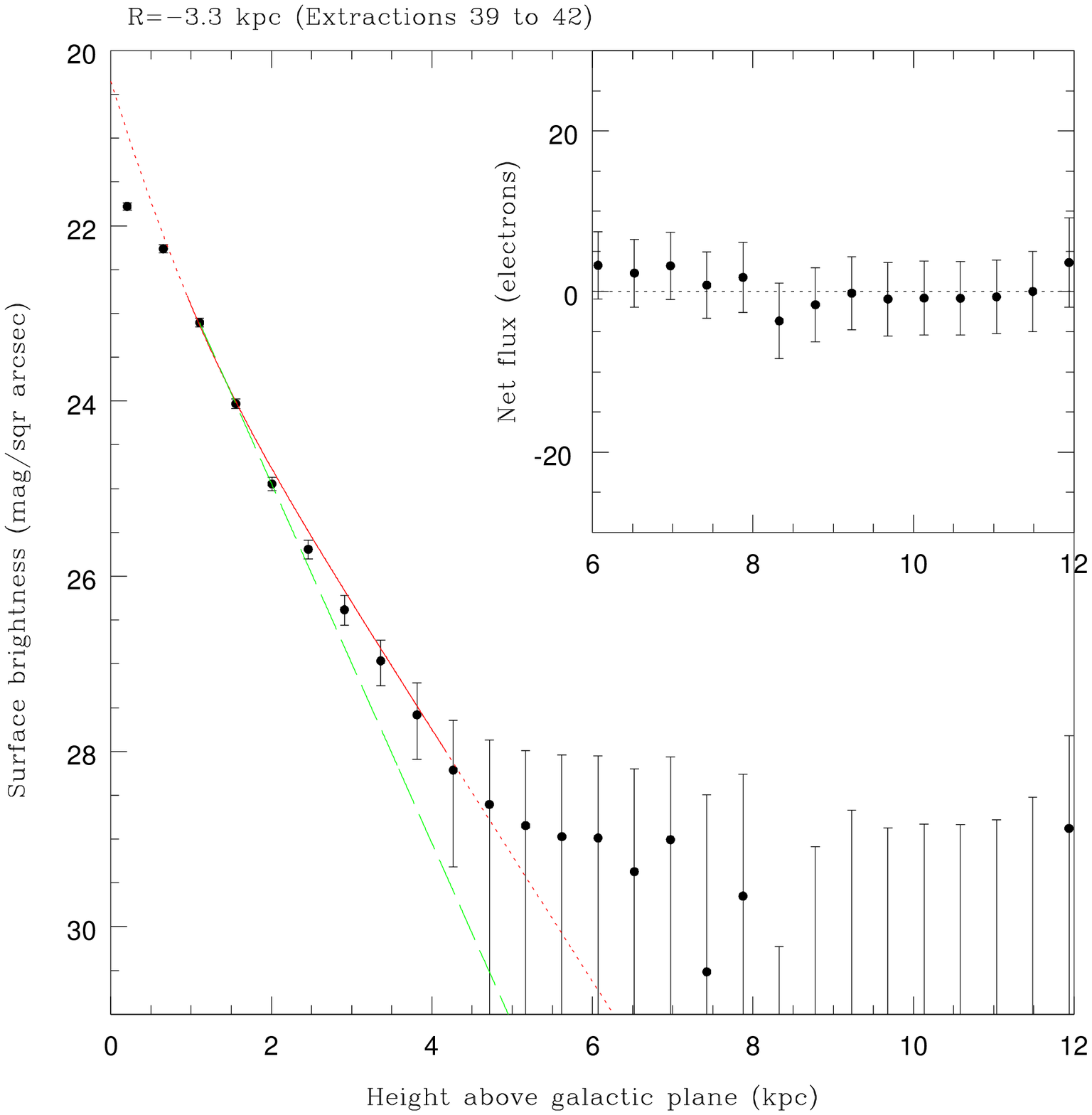,width=6.0cm}}
\hbox{\hspace{0cm}\psfig{figure=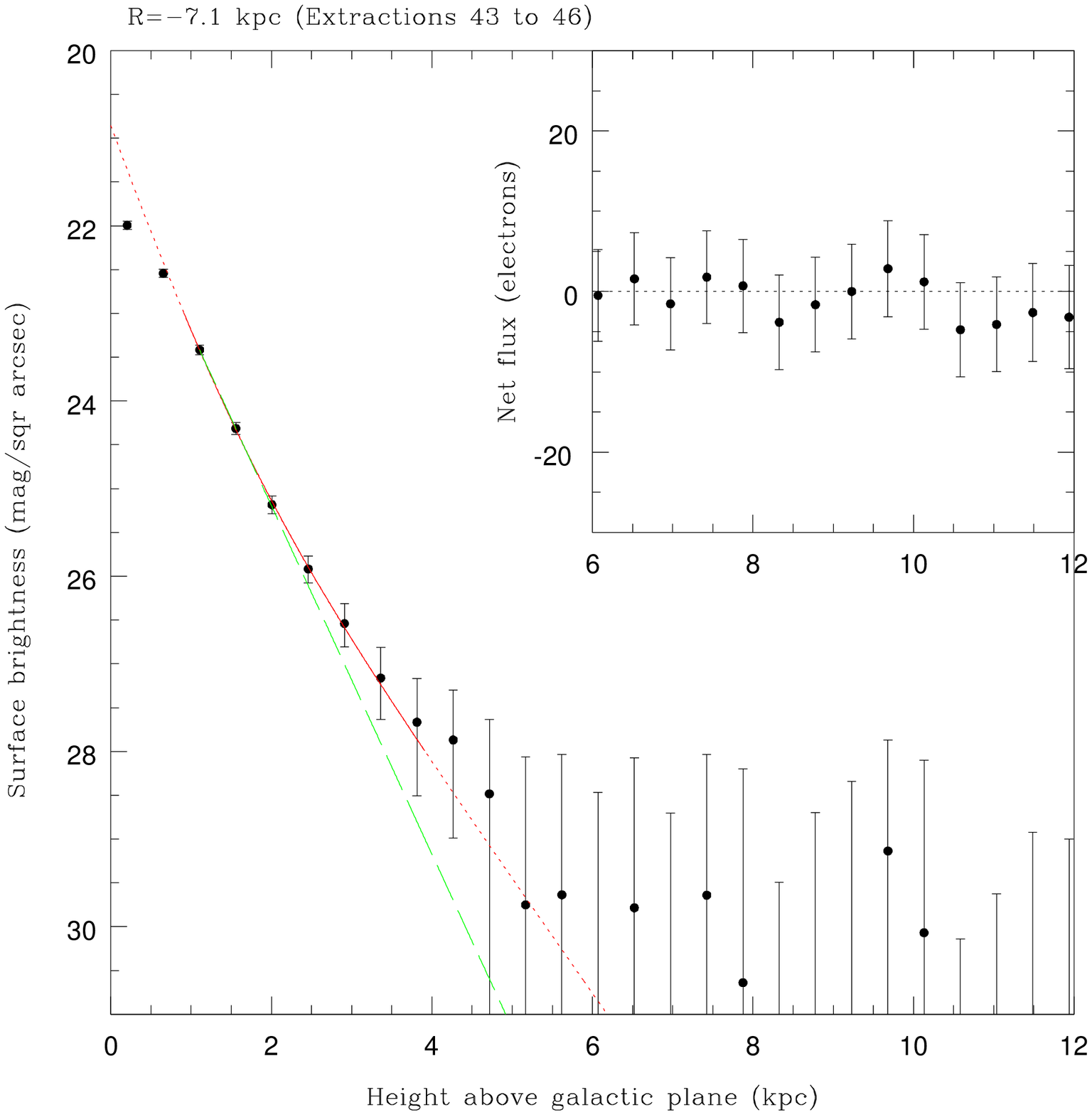,width=6.0cm}\hspace{0cm}
\psfig{figure=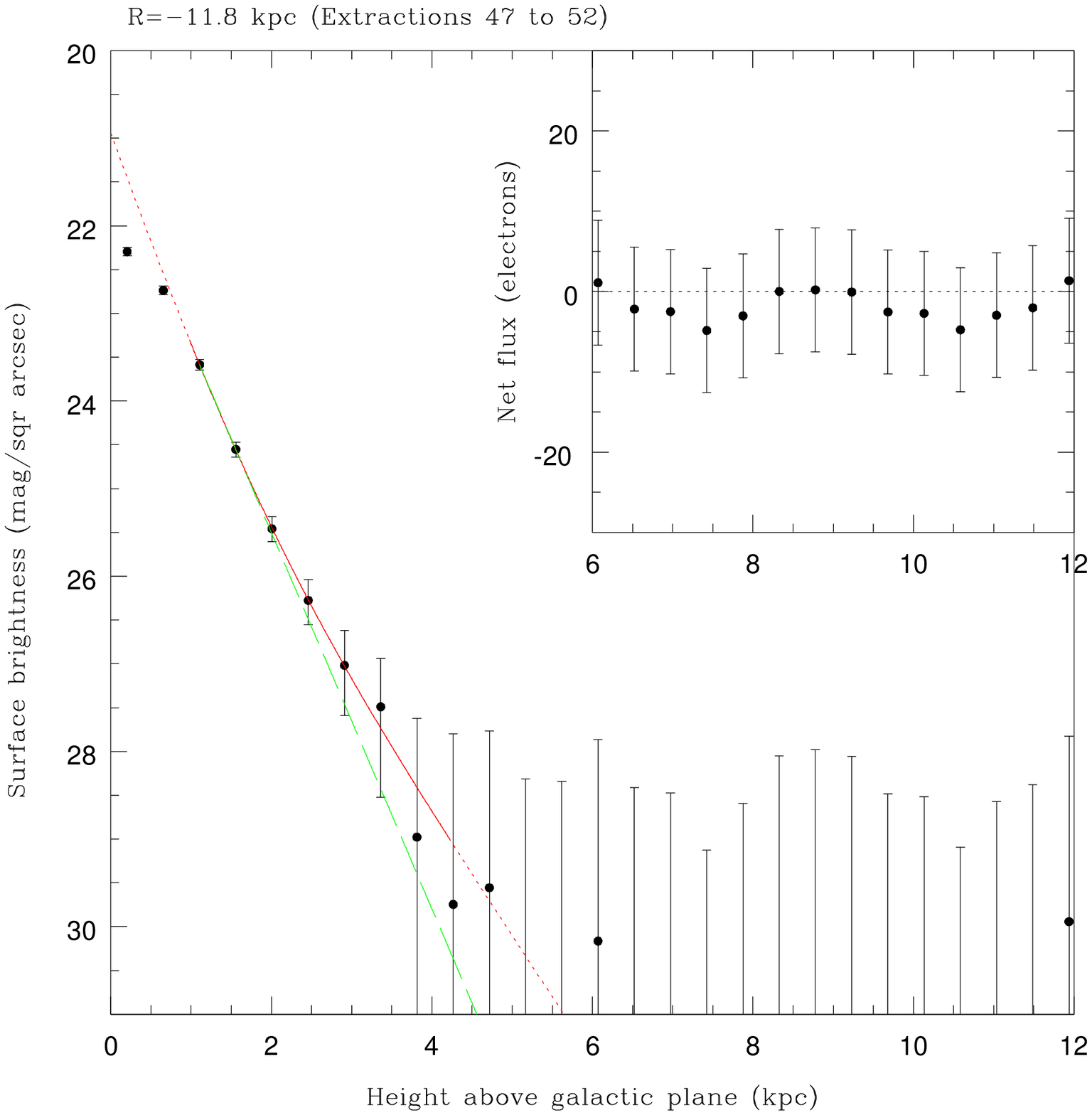,width=6.0cm}}
\vspace{0cm}
\caption{As in Fig.~\ref{fig:profilesR}, but for the
$V$-band averaged profiles through the disk of \gal.}
\label{fig:profilesV}
\end{figure*}

The systematic uncertainties in the sky level of
$\delta {\sc S}_{R} = 0.4~e^{-}~{\rm pix}^{-1}$
and $\delta {\sc S}_{V} = 0.2~e^{-}~{\rm pix}^{-1}$ correspond to errors of
only 0.0024\% and 0.0068\% per pixel in the $R$ and $V$ bands, respectively.
These systematic uncertainties are present in the
sky-subtracted profiles we present in the next section, but because they
correspond to light levels $\Delta R = 11.5$ and $\Delta V = 10.4$ magnitudes
below the sky (m$_{\rm sky}$(R) = 20.98\Msqarc\ and
m$_{\rm sky}$(V) = 21.60\Msqarc),
they are of no importance ($<$10\%) over the range of surface brightness
we consider.
The systematic uncertainty in overall calibration to a standard system
of about 5\% is relevant, but simply corresponds to a possible overall shift
in the surface brightness scale by that amount.
Note that at bright flux levels, the error in the magnitudes
is dominated by the error in the photometric conversion term, not by
$\sigma_{\rm STAT}$.   At faint flux levels, the situation is reversed.

\section{Results}  \label{sect:results}

\subsection{Analysis of the Surface Brightness Profiles}

In Figs.~\ref{fig:profilesR} and \ref{fig:profilesV},
we show the vertical extractions, averaged above and below the plane
of \gal, derived from our deep VLT imaging and discussed in Sect.
\ref{sect:profiles}.  Individual extractions
consist of the horizontal average of 21 pixel wide rectangles, with
foreground stars and background galaxies masked.  Extractions above
and below the galaxy disk were averaged about their axis of symmetry
and averaged in groups to produce the profiles shown
in the figures.
In order to display meaningfully our data at the faintest levels,
insets in Figs.~\ref{fig:profilesR} and \ref{fig:profilesV}
show fluxes on a linear scale for distances greater than 6\,kpc from the
major axis of \gal.  The scatter in each inset about zero indicates
clearly that the sky flux has been well-subtracted in our final
mosaic within our calculated uncertainties.

The deviation from pure exponential behaviour in nearly
all of the profiles indicates that presence of extended light
beyond that expected for a purely exponential stellar disk.
This motivated our choice of one- (thin disk only) and two-component
(thin+thick disks) least-squares fits to the profiles, which
are overplotted in Figs.~\ref{fig:profilesR} and \ref{fig:profilesV}
on the data.  Both components were modeled as exponential
disks (see equation~1) with the scale height $h_z$ and
central surface brightness $f_{\rm 0}$ as free parameters.
The scale heights and central surface brightnesses
(expressed in \Msqarc) derived from the simultaneous
thin plus thick disk fits are summarized in Table~\ref{tab:diskparams}.

\begin{table*}[t]
\begin{center}

\caption{Fitted Disk Structure Parameters for \gal}
\begin{tabular}{r r r r r r r r r r}
\hline
\noalign{\smallskip}
 \multicolumn{1}{c}{~}
&\multicolumn{4}{c}{Thin Disk}
&\multicolumn{1}{c} { }
&\multicolumn{4}{c}{Thick Disk}\\
 \multicolumn{1}{c}{~}
&\multicolumn{2}{c}{$\mu(0)$ (\Msqarc)}
&\multicolumn{2}{c}{$h_z$ (pc)}
&\multicolumn{1}{c}{ }
&\multicolumn{2}{c}{$\mu(0)$ (\Msqarc)}
&\multicolumn{2}{c}{$h_z$ (pc)}\\
 \multicolumn{1}{c}{R (kpc)}
&\multicolumn{1}{c}{$R$}
&\multicolumn{1}{c}{$V$}
&\multicolumn{1}{c}{$R$}
&\multicolumn{1}{c}{$V$}
&\multicolumn{1}{c}{}
&\multicolumn{1}{c}{$R$}
&\multicolumn{1}{c}{$V$}
&\multicolumn{1}{c}{$R$}
&\multicolumn{1}{c}{$V$} \\
\hline\hline

$-$17.5	& 24.2$^{+1.3}_{-1.2}$   & 22.7$^{+1.5}_{-1.3}$   & 340$\pm$240 & 350$\pm$210 & & 22.49$^{+0.90}_{-0.85}$ & 22.9$^{+1.8}_{-1.5}$     & 600$\pm$ 120 & 600$\pm$230 \\
$-$11.8	& 22.3$^{+1.3}_{-1.2}$   & 22.1$^{+1.4}_{-1.2}$   & 520$\pm$190 & 320$\pm$210 & & 23.0$^{+1.9}_{-1.8}$    & 22.08$^{+1.0}_{-0.96}$   & 720$\pm$ 180 & 600$\pm$160 \\
$-$7.1	& 20.83$^{+0.03}_{-0.03}$& 21.36$^{+0.13}_{-0.12}$& 421$\pm$18  & 410$\pm$140 & & 23.05$^{+0.34}_{-0.26}$ & 23.0$^{+1.1}_{-1.1}$     & 870$\pm$ 55  & 760$\pm$270 \\
$-$3.3	& 21.25$^{+0.12}_{-0.10}$& 21.37$^{+0.09}_{-0.09}$& 494$\pm$35  & 428$\pm$91  & & 22.43$^{+0.59}_{-0.38}$ & 22.73$^{+1.6}_{-0.62}$   & 807$\pm$ 59  & 820$\pm$140 \\
0	& 20.01$^{+0.04}_{-0.04}$& 20.49$^{+0.17}_{-0.17}$& 346$\pm$9   & 360$\pm$46  & & 22.11$^{+0.09}_{-0.08}$ & 22.53$^{+0.57}_{-0.43}$  & 808$\pm$ 16  & 806$\pm$84 \\
3.3	& 20.19$^{+0.02}_{-0.02}$& 20.68$^{+0.12}_{-0.10}$& 396$\pm$13  & 384$\pm$75  & & 22.33$^{+0.22}_{-0.18}$ & 22.29$^{+1.7}_{-0.63}$   & 806$\pm$ 33  & 740$\pm$120 \\
7.1	& 20.61$^{+0.04}_{-0.03}$& 21.03$^{+0.14}_{-0.12}$& 416$\pm$22  & 422$\pm$87  & & 22.58$^{+0.33}_{-0.25}$ & 23.21$^{+0.90}_{-0.88}$  & 856$\pm$ 52  & 880$\pm$260 \\
11.8	& 20.90$^{+0.06}_{-0.06}$& 21.05$^{+0.28}_{-0.22}$& 431$\pm$33  & 390$\pm$110 & & 23.62$^{+1.1}_{-0.55}$  & 23.0$^{+1.5}_{-1.4}$     & 950$\pm$ 160 & 760$\pm$320 \\
17.5	& 21.44$^{+0.51}_{-0.34}$&                        & 324$\pm$51  &           &~~~& 24.85$^{+0.90}_{-0.49}$ &                          & 1100$\pm$ 220&              \\

\noalign{\smallskip}
\hline
\label{tab:diskparams}
\end{tabular}
\end{center}
\vspace*{-0.6cm}
\noindent{ }
{\footnotesize
Note: these structure parameters are the results of the model fits to 
the observed profiles and have not been corrected for projection effects.}
\noindent{ }
\end{table*}

\subsection{Could the Faint Extended Emission be Scattered Light?} \label{sect:psf}

The first-order effect of turbulence in the atmosphere  
causes the radial point spread functions (PSFs) of point-like objects
measured by an astronomical detector to have a roughly gaussian shape,
but many effects, including scattering in the telescope optics, 
can lead to broader wings.  Although the simple optics
of the VLT test camera (\cite{giacconi99})
should minimize such a scattering, faint wings in the PSF are still present.
To quantify the effect of these wings on our
faint surface brightness photometry, we measured the PSFs of
isolated fainter stars in the field of \gal\ and
bright standard stars observed on the same nights as our science frames.

In order to be meaningful, such PSFs must be
constructed with high signal-to-noise data.
A conservative estimate of the precision
required can be made by assuming that all of the light of \gal\ is confined 
to a point at a distance equal to the angular separation between the 
center of the galaxy and the most distant point above the plane we consider. 
Such an estimate shows that ,at 6~kpc above the plane, the amount of 
$R$-band light brighter than 30~\Msqarc\ scattered from \gal\
can be quantified easily if the PSF is known to a precision of
$\sim 2.5 \times 10^{-6}$ at that distance.

Only three relatively isolated stars near the center of the \gal\ mosaic 
are available; the $R$- and $V$-band images of these were masked, added, 
and azimuthally averaged.  
The result is shown in Fig.~\ref{fig:psfs}.  Note that although the
extended emission around \gal\ is seen more clearly in the $R$-band,
median seeing in $R$ is better than in $V$.
At faint light levels, both the $R$ and $V$ faint-star
PSFs have broader wings than a pure gaussian.
Unfortunately, the statistical noise in these faint-star PSFs,
and the relative size of the systematic photometric uncertainties
over the relevant radii, precludes measurement in the wings
to the accuracy we require.  In principle, saturated stars on
the mosaic could be used to study the wings of the PSF,
but our small field contained only two;
one has a near bright neighbor and the other does not fall
on the $V$-band mosaic.

\begin{figure}
\vspace{0cm}
\hspace{-0.4cm}\epsfxsize=9.3cm \epsfbox{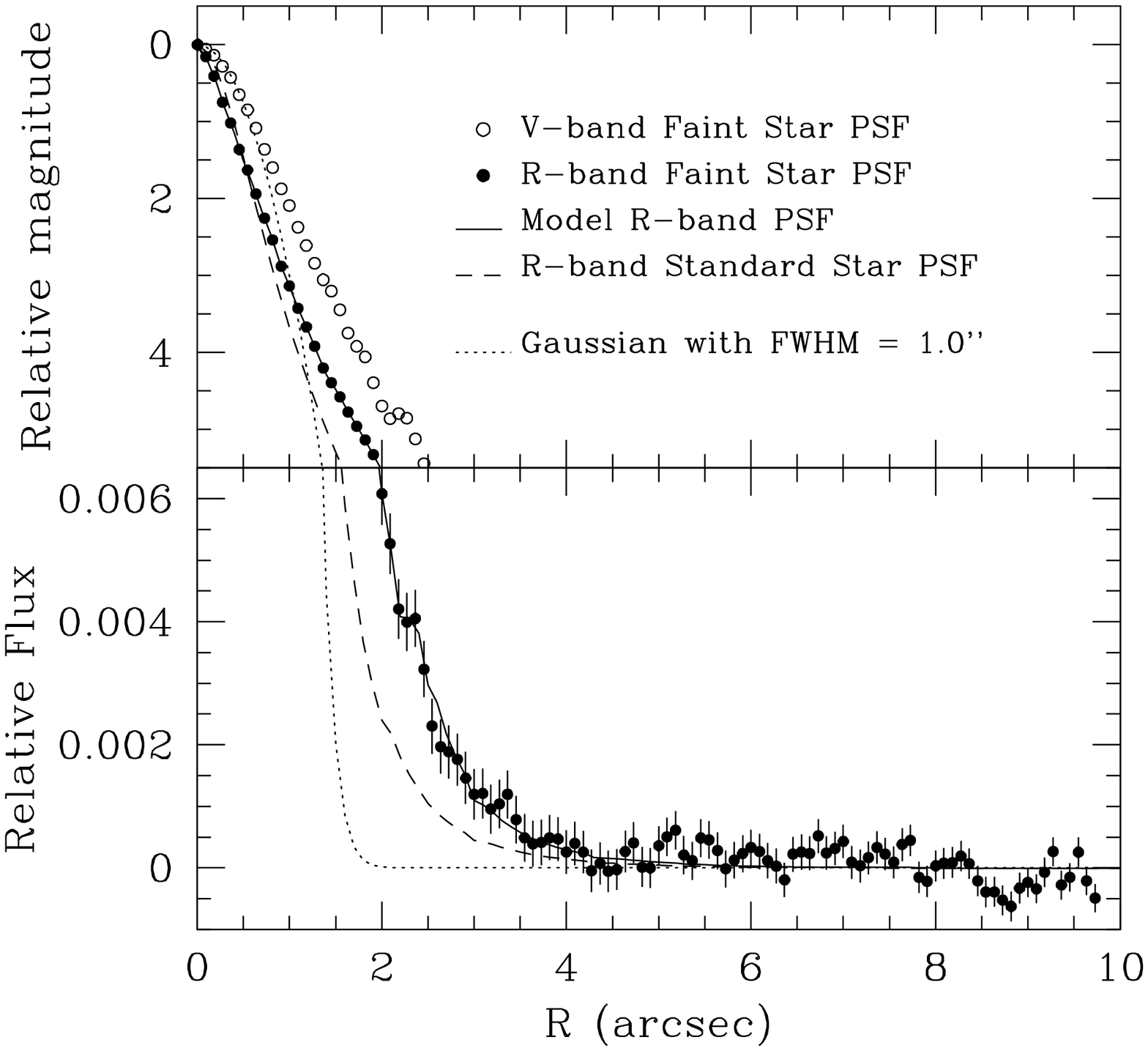}
\vskip -0.8cm
\caption{
Measured radial point spread functions (PSF) for standard stars and 
fainter stars in the \gal\ field shown on magnitude (top)
and linear (bottom) scales chosen to display the full dynamic range
in a meaningful way.
The data shown in the two plots
are otherwise identical, and the apparent change in PSF shape in
the linear plot is solely due to the different scales of the logarithmic
and linear representation.
The $R$ and $V$ PSFs taken from
isolated stars in the \gal\ field
are shown as solid and open dots, respectively.  
The higher signal-to-noise PSF derived from a brighter standard star
is shown as the dashed line.
The model PSF formed by matching the faint star PSF to the
wings of the bright star PSF is indicated by the solid line.
For comparison, a gaussian with FWHM = 1\arcsec\ (dotted line)
is also shown.
Error bars shown in the lower panel represent photon noise only.
}
\label{fig:psfs}
\end{figure}

We therefore study the PSF wings using much brighter
standard stars imaged during the same observing run.  We build a model 
PSF directly from the data, using the isolated three stars on the 
mosaic of \gal\ to derive the PSF out to 2.5\arcsec, and a bright 
reference star observed with similar seeing to derive the PSF
from 2.5 to 16\arcsec\
(i.e., out to 8~kpc above the galaxy plane).
Since the seeing was slightly
better during the imaging of the reference star, the
standard star profile was horizontally displaced to create a smooth match to the inner
PSF derived from the mosaic.  The result is shown as the thin
solid line in Fig.~\ref{fig:psfs}.
The $R$-band PSF of the standard star is consistent with zero at the level
of $2.5 \times 10^{-6}$ from 7 to 12 arcsec ($\sim 3.5$ to 6~kpc),
satisfying the conservative requirement that we derived above.
We conclude, therefore, that the PSF is well enough understood
to determine its effect on the observed shape of
the vertical surface brightness profiles of \gal.

\begin{figure}
\vspace{0cm}
\hspace{0cm}\epsfxsize=9cm \epsfbox{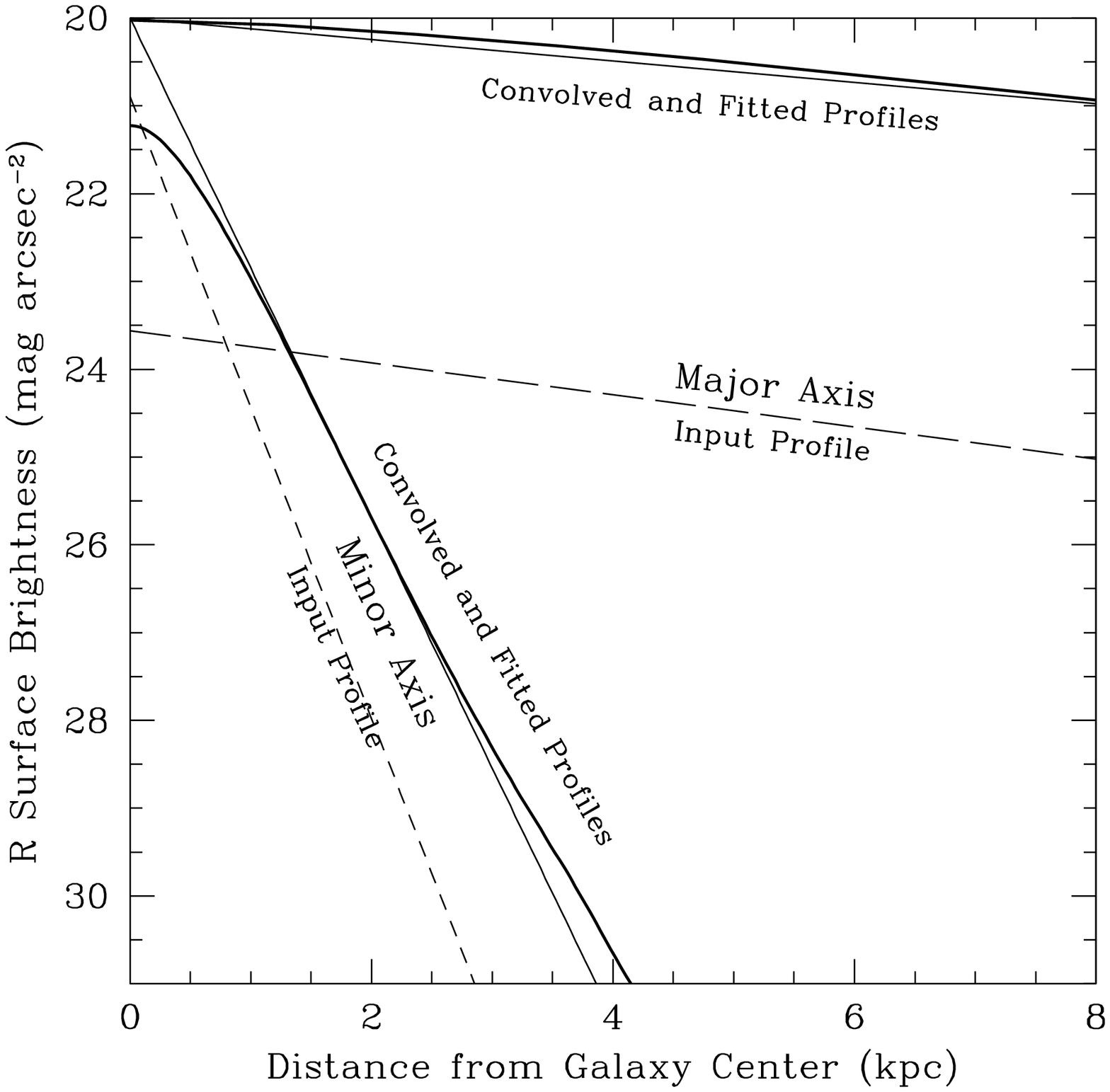}
\vskip -0.2cm
\caption{The effect of inclination and PSF convolution
on the observed radial and vertical surface brightness
profiles along the major and minor axes of a thin exponential disk
similar to that of \gal.  The inferred
intrinsic vertical and face-on radial profiles are shown as thin
dashed lines.  The thick solid line indicates the
result after inclination by 88 degrees and convolution
with the high signal-to-noise PSF
determined from isolated faint stars
on the science mosaic and a bright standard star.
The vertical (minor axis) and radial (major axis) profiles
of an exponential thin disk with typical
fitted parameters for the projected scale height, $h_z = 380~$pc,
and scale length, $h_R = 8.9~$kpc,
(see Sect.~\ref{sect:disks}) are shown as thin solid lines.
}
\label{fig:psfconvolve}
\end{figure}

In order to examine whether the extended light apparent in
Figs.~\ref{fig:profilesR} and \ref{fig:profilesV} might be
due to thin disk light scattered through the
broad wings of the PSF to other positions on the detector,
we convolved a model exponential disk with intrinsic
structural characteristics similar to those of \gal\ with our model
$R$-band PSF.
The intrinsic thin disk model parameters reported in
Sect.~\ref{sect:disks}
were determined by requiring that, after inclination and convolution with
the observed PSF, the projected thin-disk fitted parameters were retrieved.
The degree to which the thin disk fits are reproduced is illustrated 
in Fig.~\ref{fig:psfconvolve}.

Due to its high inclination, \gal\ has an observed surface brightness along
its length that is much larger than the intrinsic (input) face-on
value.
Except for the central regions, which suffer a net loss of light from
scattering, the primary effect of inclination --- and to
a lesser extent scattering --- is to increase the amount of light observed
at a given angular distance from the plane of \gal.  The result (output)
is an observed profile that is approximately exponential, but
with a projected scale height larger than the intrinsic value.
More importantly, however, Fig.~\ref{fig:psfconvolve} clearly
illustrates that for surface brightnesses brighter than
$R \approx 28.5~$\Msqarc,
no substantial light in excess of the projected thin disk
profile is generated by inclination and scattered light effects.
The extended light $R > 26.5~$\Msqarc\ in many of the profiles
of Fig.~\ref{fig:profilesR}, therefore, must have another cause;
we conclude that it is intrinsic to the galaxy itself.
This conclusion is supported by the constant color (or possible
slight reddening) of the extended light with increasing distance
from the galaxy plane, despite the fact that the
scattering in the $V$-band images is larger than that in $R$
as measured from the stellar PSF on the science mosaic.

\section{The Thin and Thick Disks of \gal} \label{sect:disks}

The extended light in \gal\ is reasonably well
fit by a thick exponential disk with nearly constant projected scale
height ($h_z$) as a function of galactocentric radius (R) along
the major axis of the galaxy.  This is illustrated in
Figs.~\ref{fig:profilesR} and \ref{fig:profilesV}, and in
Fig.~\ref{fig:diskhzs}, where the fitted values of $h_z$
for each averaged extraction are shown in both the $R$ and $V$
bands for the thick and thin exponential disk components.
The error-weighted mean of the projected scale heights are:
$h_z^{\rm thin} = 380 \pm 35~$pc and $h_z^{\rm thick} = 810 \pm 40~$pc in the
$R$-band and
$h_z^{\rm thin} = 380 \pm 45~$pc and $h_z^{\rm thick} = 760 \pm 75~$pc in the
$V$-band.
The projected scale length, $h_R$, of the thin disk
is more difficult to assess, but
is estimated from the fitted values of $\mu(0)$ as a function of
position along the major axis to be about $8.9 \pm 1.5~$kpc in
both bands.
The projected scale length of the
fitted thin disk is indeterminate from the V-band frames, but is
consistent, within the uncertainties, with the projected scale length of the
thin disk in the R-band.

\begin{figure}
\vspace{0cm}
\centerline{
\hspace{0cm}\epsfxsize=7.5cm \epsfbox{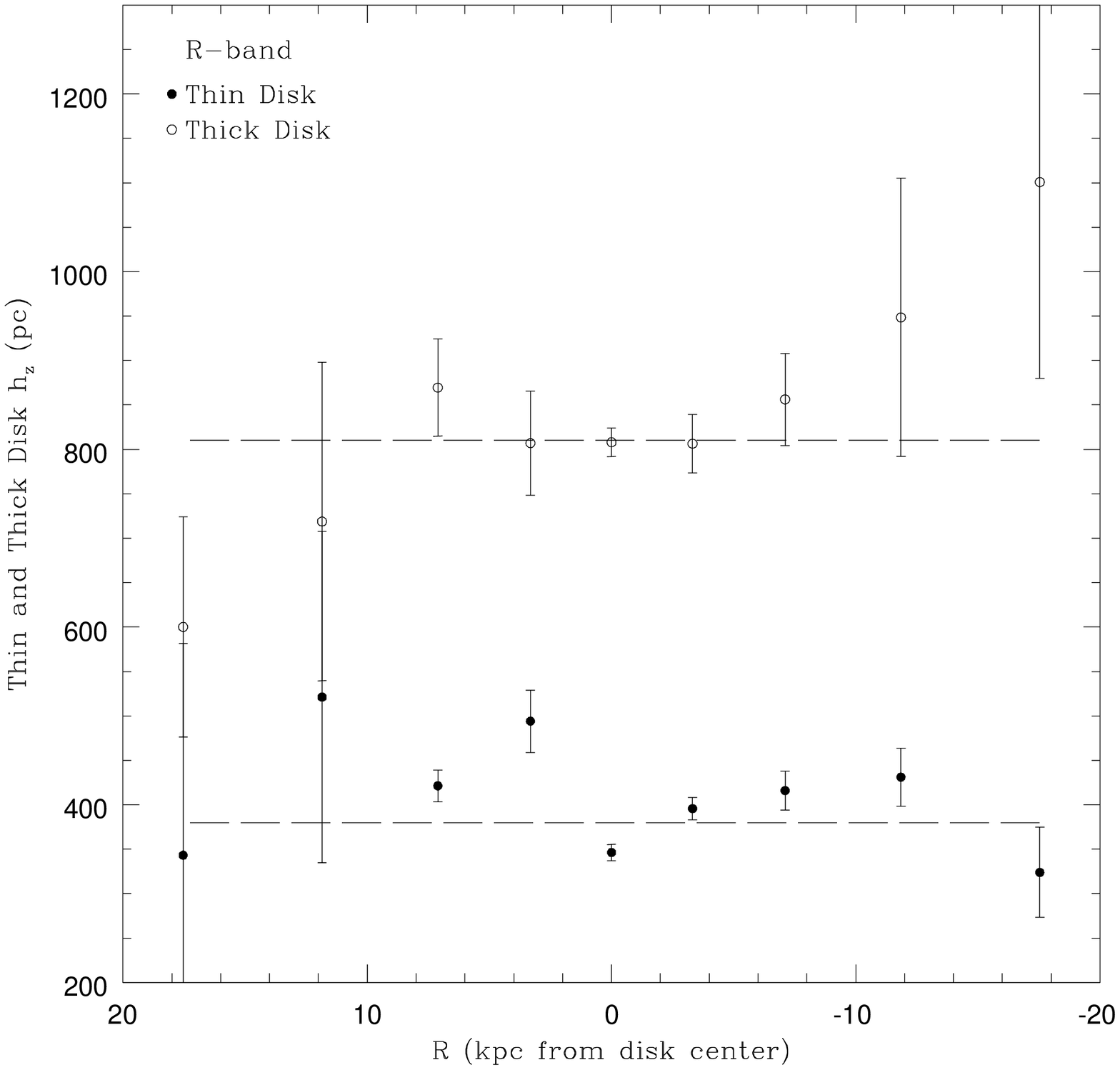}
}
\centerline{
\hspace{0cm}\epsfxsize=7.5cm \epsfbox{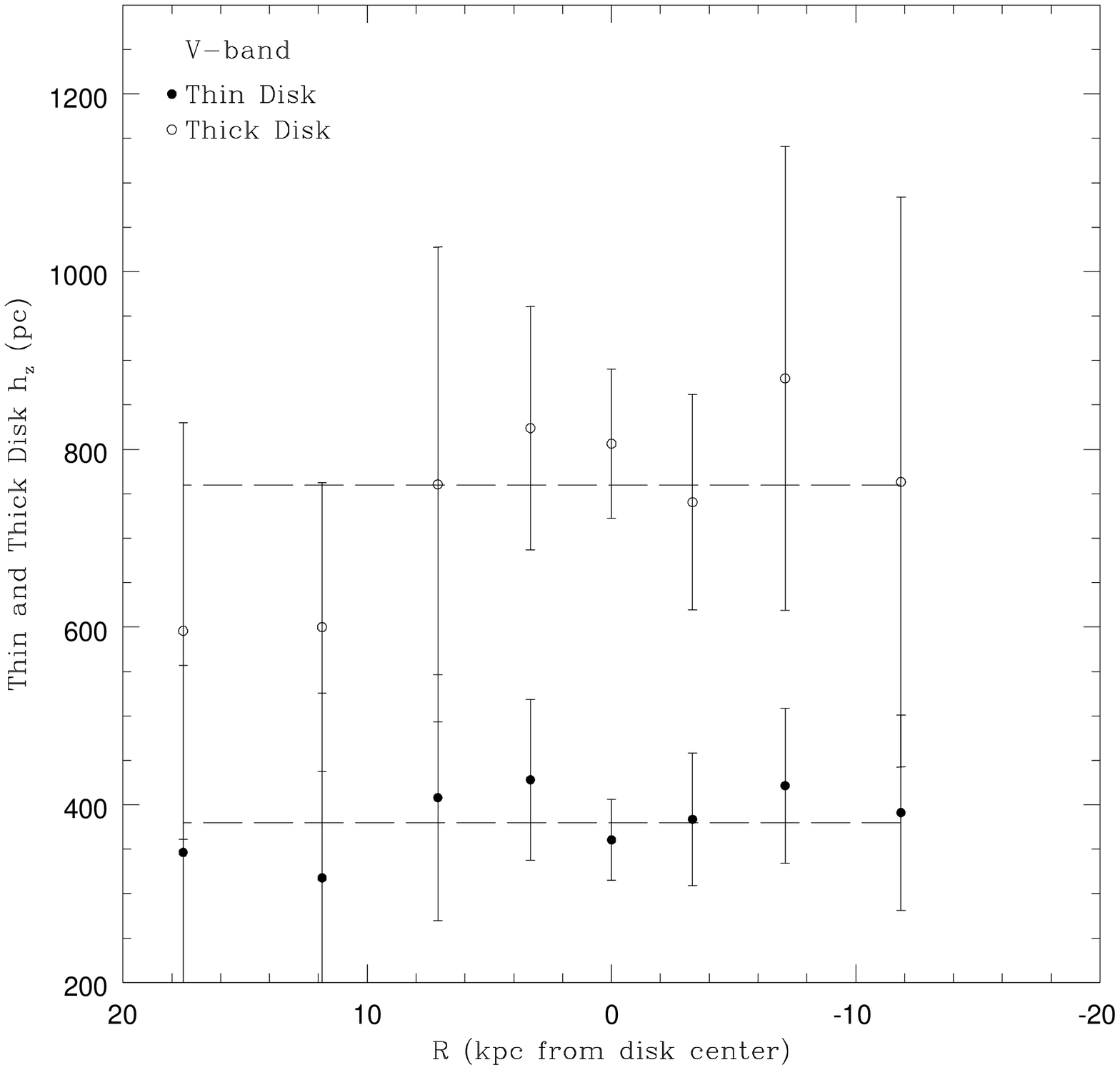}
}
\vskip -0.2cm
\caption{The fitted values of $h_z$ for two-component (thin +
thick exponential disks) model of the vertical surface brightness
extractions of \gal\ in both $R$ (top) and $V$ (bottom) bands.
The error bars indicate the formal errors of the fit, and
are clearly larger in $V$-band and at larger galactocentric radius
R where the $S/N$ is poorest.  Horizontal dashed lines indicate the
error-weighted mean of $h_z$ for the two components in each
photometric band.
}
\label{fig:diskhzs}
\end{figure}

\begin{figure}
\vspace{0cm}
\centerline{
\hbox{\hspace{0cm}\psfig{figure=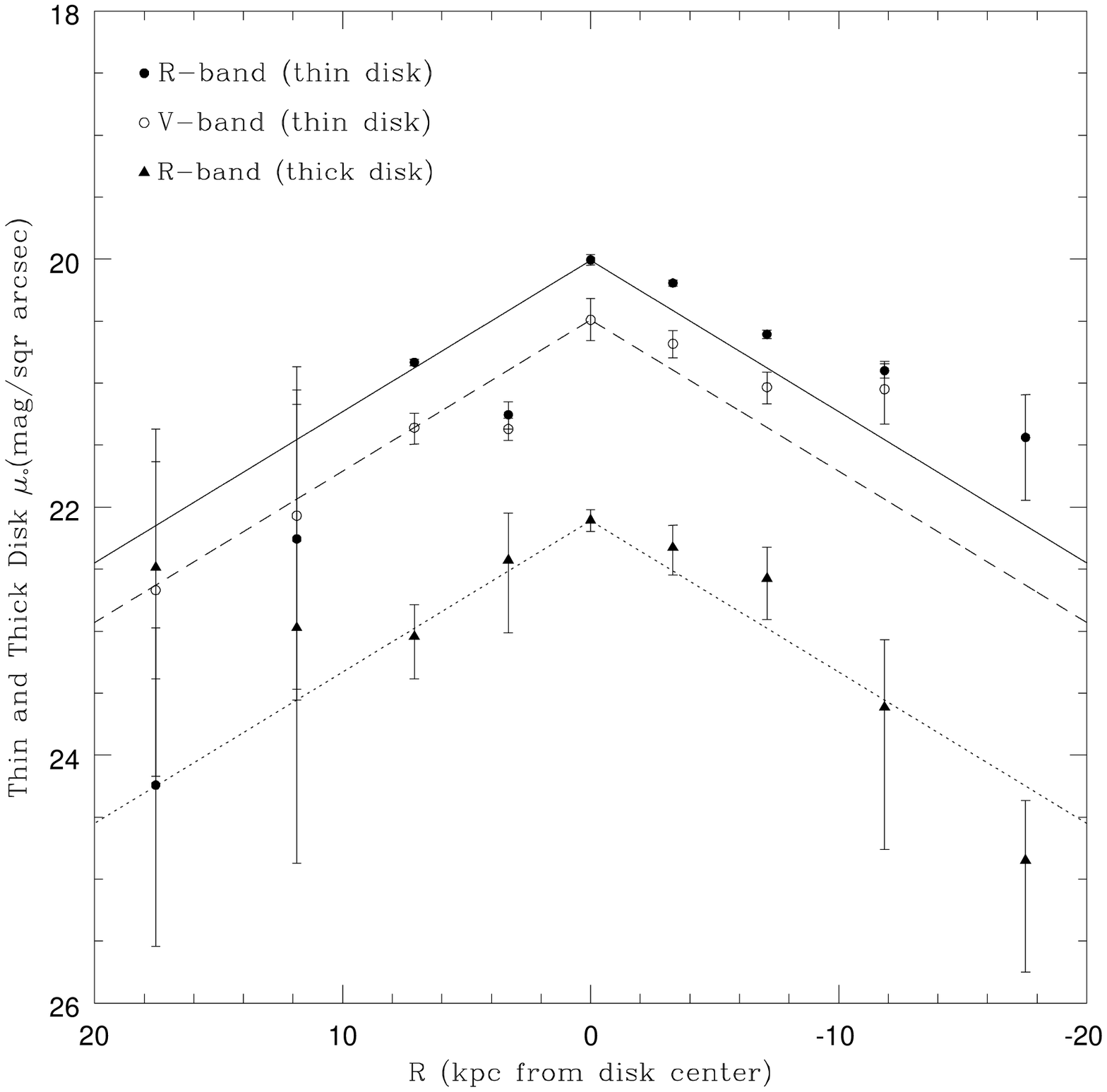,width=8.3cm}\hspace{0cm}}}
\vskip -0.2cm
\caption{The fitted values of $\mu$(0) for our two-component
fits to the vertical surface brightness extractions of \gal.
in both $R$ and $V$ bands.  The $V$-band thick disk fit is
not shown due to its poorer $S/N$.
As a rough guide, an exponential of the form $f(R) = f(0) * exp(-R/h_R)$
has been drawn on the data using a constant scale length
of $h_R = 8.9~$kpc and the peak central surface brightness ($\mu(0)$) of
each component.  The intent is to show that a $h_R > 8$~kpc is consistent
with the data, and that the $h_R$ is similar for both thin and thick disks.
}
\label{fig:diskmu}
\end{figure}

When deprojected and deconvolved (that is,
taking into account line-of-sight effects due to the
inclination of the galaxy and seeing), the true
face-on surface brightness of the thin disk in the $R$-band is
$\mu_{R,0}^{\rm thin} = 23.6~$\Msqarc,
with a true scale height and scale length
of ${\rm h_{z,0}^{\rm thin}} = 310 \, $pc
and ${\rm h_{R,0}^{\rm thin}} = 5.9\, $kpc, respectively.
These estimates were made by convolving model thin exponential disks
inclined at 88$^o$ with the measured PSF, and requiring that
the resulting vertical and radial profiles matched those fitted to
the observed profiles
(see fig~\ref{fig:psfconvolve}).
The intrinsic thin disk scale heights and lengths in the $V$-band are
the same as those in the $R$-band, within uncertainties.
The projected scale height, $h_z$,
is larger than the intrinsic value $h_{z,0}$
primarily due to convolution with the comparably sized PSF.
On the other hand, the projected scale length, $h_R$, is larger
than $h_{R,0}$ because of line-of-sight effects due to the
extreme inclination of the galaxy.
The thin disk has an inferred face-on surface brightness in $V$
of $\mu_{V,0}^{\rm thin} = 24.1~$\Msqarc, implying an
intrinsic color of $V - R \approx 0.5$ for the thin disk.
Since the color is found by extrapolating the fitted
parameters into the plane of the disk, it is relatively, though
not completely, insensitive to dust and
clumpy luminosity from HII regions.
We estimate that the uncertainty in our inferred intrinsic parameters
is about $10-15$\%, primarily coming from uncertainties in
inclination and the fit parameters.

The structural parameters of the
extended light are more uncertain, but also
much less affected by inclination
and seeing effects.  We have not attempted, therefore, to
deproject the thick disk scale parameters, but expect that in
the $R$ band the intrinsic scale height is
close to the projected value of $h_z^{\rm thick} = 810 \pm 40~$pc,
while the true scale length of the thick disk is between $\sim$6 and 9~kpc,
(the intrinsic thin disk and projected thick disk values, respectively).
The value of the central surface brightness of the thick disk
is uncertain, but can be constrained.
For a pure exponential disk, the edge-on
central surface brightness $\Sigma_{\rm EDGE}(0)$ (in linear units)
can be shown to be given by
$\Sigma_{\rm EDGE}(0) = ({\rm h_{R,0} / h_{z,0}}) \Sigma_0(0)$,
where $\Sigma_0(0)$ is the face-on central surface brightness.
If we assume that the fitted value $\mu_{R}^{\rm thick} = 22.1~$\Msqarc\
of the thick disk in the $R$-band is a good approximation to the
actual edge-on value for the
thick disk, then, based on our estimates of these quantities and
their uncertainties, we can deduce that
$24.1 < \mu_{R,0}^{\rm thick} < 24.9~$\Msqarc.
The PSF may have a small effect that would cause the
fitted value to be higher than the actual value, in which
case these constraints would be pushed to slightly fainter magnitudes.
The detection of the thick disk in the $V$-band
is less secure,
both because the $S/N$ of our relative surface brightness photometry
is lower in $V$ and because the PSF (and thus scattered light
problems) is larger in $V$.  Furthermore, beyond galactocentric
radii of 5~kpc, there is only a small statistical difference
in the inferred scale heights of the fitted thin and thick disks
(Fig.~\ref{fig:diskhzs}), and the extrapolated in-plane surface
brightness of the thick component in the $V$-band
shows no clear trend with major axis radius.

The intrinsic $R$-band scale heights of the thin and thick disk
components of \gal\ are similar to those of the Milky Way,
but because the intrinsic scale length of its thin disk
is larger than the commonly accepted Galactic value of
$h_{R,0} = 3 - 3.5~$kpc (see references in \cite{sackett97}),
the ratio ${\rm h_{R,0} / h_{z,0}}$ is
$\sim$50\% larger for \gal\ than for the Galaxy.
Since the total luminosity of any pure exponential disk is given by
$L = 2 \pi \Sigma_0(0) h^2_{R,0}$, if the intrinsic
scale lengths of the thick and thin components are equal,
the ratio of total light in each is given by the
ratio of their intrinsic central surface brightness.
Together with the constraints on $\mu_{R,0}$ for the two components derived
above, this assumption implies that the thick disk
contributes $\sim$20-40\% of the total $R$-band light of \gal, excluding
the light in individual masked HII regions.

Finally, we note that these constraints on the luminosity contribution
of the thick disk imply
a combined (thin+thick disk) face-on central surface brightness
for \gal\ of $\mu_{R,0} > 23.1$.
Since the $B - R$ color of the galaxy is certainly greater
than zero, and probably $\ge 0.5$, this places \gal\ firmly
in the class of low surface brightness (LSB) galaxies,
which are generally defined as those
disks with $B$-band face-on central surface brightnesses
$\mu_{B,0} > 23$\Msqarc\ (c.f., \cite{deblok95}).

\section{Summary and Conclusions}

We have used the VLT test camera on UT1 to obtain deep surface
brightness photometry of the edge-on LSB galaxy \gal\ in the $V$
and $R$-bands.  Careful masking of foreground and background
objects to obtain an accurate value of the sky flux
on our science mosaics, and an analysis of flat-fielding
uncertainties -- both statistical and systematic -- on a variety
of spatial scales, allow us to estimate
confidently the total uncertainty in our deep
surface photometry.  We conclude that on the size scales
important for probing faint, extended structure, we reach
$V = 28$ and $R = 29~$\Msqarc.  A detailed analysis of the
PSF of the images, derived from faint isolated stars on the
mosaic and standard stars, indicates that scattered light
affects the extended vertical luminosity profiles of \gal\
only for $R > 28.5~$\Msqarc.

Extended light in excess of that expected for a single-component
thin disk is detected at about $R > 26.5~$\Msqarc\ in nearly
all vertical profiles perpendicular to and up to
17~kpc along the major axis of \gal.
The same component may have also been detected in the $V$ band
frames, but the lower $S/N$ of these frames and the larger
PSF in $V$ make this detection less robust.   Given the geometric
form of the extended light in this apparently bulgeless galaxy,
we interpret the faint $R$-band light as a thick disk.

Two-component exponential disk fits were made
to the observed surface brightness profiles and used to
determine projected and -- after deprojection and
deconvolution -- intrinsic structure parameters for
the thin disk of \gal\ in the $V$- and $R$-bands and for the thick
component in $R$. In particular, we find:

\begin{itemize}

\item The thin disk has projected scale heights perpendicular
        to the major axis of $h_z^{\rm thin} = 380 \pm 35~$pc
        in the $R$-band and $h_z^{\rm thin} = 380 \pm 45~$pc in $V$.
        The projected scale length of the thin disk is
        $8.9 \pm 1.5~$kpc in both bands.

\item   After deprojection and deconvolution with the PSF derived
        from our observations, we estimate that, within the errors,
        the intrinsic thin disk scale heights are 80\% of the
        measured values.  The intrinsic scale length of the
        thin disk is $2/3$ of the fitted value.

\item   The face-on central surface brightness of the thin
        disk is estimated to be  $\mu_{0}^{\rm thin} = 23.6~$\Msqarc\
        in $R$ and $\mu_{0}^{\rm thin} = 24.1~$\Msqarc\ in $V$.

\item   The thick disk, which is detected robustly in our $R$
        surface photometry, has a projected scale height of
        $h_z^{\rm thick} = 810 \pm 40~$pc in the $R$-band and
        $h_z^{\rm thick} = 760 \pm 75~$pc in $V$.
        The projected scale length of the thick disk
        cannot be determined precisely from our observations,
        but within uncertainties is consistent with that of the thin disk.

\item   The intrinsic scale parameters of the thick disk
        are somewhat smaller than these measured projected ones,
        though not as dramatically different as the
        projected and intrinsic parameters of the thin disk.

\item   Simple considerations lead to an estimate for the
        face-on central surface brightness of the thick
        disk of  $24.1 < \mu_{R,0}^{\rm thick} < 24.9~$\Msqarc.

\item   The thick disk is likely to contribute 20-40\% of the
        total (old) disk $R$-band luminosity of the galaxy.

\item   The total central surface brightness of
        the (thin + thick) disk is $\mu_{R,0} > 23.1~$\Msqarc,
        which places \gal\ securely in the category of
        low surface brightness (LSB) galaxies.

\end{itemize}

This detection of a thick disk adds to only a few others known
in external galaxies (see Sect.~1), and to our knowledge is the
first known thick disk in an LSB galaxy.  The thick and thin disks
of \gal\ have similar scale heights as their corresponding
components in the Milky Way, but larger scale lengths.
Importantly, the thick disk of \gal\ appears to
contribute a larger fraction of the
overall old disk light than does the Galactic thick disk.
(Young HII regions have been masked
and so do not enter into the extrapolated estimates we have made.)
A prominent thick disk in \gal\ is particularly interesting since,
compared to their high surface brightness cousins,
LSB galaxies are thought to be more dark-matter dominated
and to have less evolved disks.
The VLT observations reported here suggest that,
at least in the case of \gal,
such an unevolved thin disk can coexist with a substantial
thick luminous component, perhaps providing a clue to the
formation mechanism of thick disks in all spirals.

\begin{acknowledgements}

We are grateful to the ESO VLT Science Verification team for
their assistance in obtaining the data analyzed here and to
Edwin Valentijn for useful discussions.
MJN acknowledges support by the European Commission, TMR Programme, Research 
Network Contract ERBFMRXCT96-0034 ``CERES.''  PDS thanks
the Anglo-Australian Observatory, Epping and the Institute for
Advanced Study, Princeton for hospitality during the completion of
some of the work presented here.
This research has made use of the NASA/IPAC Extragalactic Database (NED) 
which is operated by the Jet Propulsion Laboratory, California Institute of 
Technology, under contract with the National Aeronautics and Space 
Administration. 

\end{acknowledgements}

\newpage

\begin{appendix} 

\section{Photometric Uncertainties}

Accurate detection and characterization of faint surface brightness 
features in galaxies requires a thorough understanding of the 
uncertainties in CCD photometry at very low light levels.
Both systematic and statistical uncertainties are present, and 
can affect the photometry over different spatial scales.
We consider here seven different sources of photometric uncertainty,
and combine them to create an error budget for an area of arbitrary 
size in the combined mosaic of the \gal\ field.  The error bars presented in 
Figs.\ref{fig:profilesR} and \ref{fig:profilesV}
are calculated according to this error budget.

Our analysis of photometric uncertainties must reflect the 
process by which the deep, masked mosaics from which we derive 
surface brightness profiles were generated.  
In what follows, all fluxes and uncertainties 
will be expressed as numbers of electrons $e^-$, and $i$ will be used 
as an index to label one of the $N_f$ individual frames that were 
co-added to form the mosaic at that position. 

The flux $F_{\rm\sc SM} (x,y)$ at any sky position in the mosaic is defined by:

$$
F_{\rm\sc SM} (x,y) = \left( \frac{\zeta}{\sum_i^{N_f} {\cal S}_i} \right) \, 
\sum_i^{N_f} e^-_i(x,y)
$$
$$
~~~~~ \equiv {\rm C_{\sc SM}} \, \sum_i^{N_f} e^-_i(x,y)  ~~,
$$

\noindent 
where ${\cal S}_i$ is the sky value (in $e^-$) on frame $i$, 
that is, the median of all background (non-object, non-masked) pixels 
in frame $i$. 
The quantity $\zeta$ is a normalization constant defined by  

$$
\zeta \equiv 
{ \frac{1}{N_f} \sum_i^{N_f} 
\left[ \frac{1}{N_{sky}}\sum_j^{N_{sky}} e^-_{ij} \right] }
$$

\noindent 
where $j$ runs over all sky pixels on frame $i$, and $i$ runs over 
all frames used anywhere in the matrix.  On the other hand, note   
that if frame $i$ did not contribute to the flux 
at position $(x,y)$, due to it being masked, then $e^-_i(x,y) = 0$ 
and ${\cal S} _i = 0$. 
Thus ${\rm C_{\sc SM}}$ is a property of submosaics --
portions of the mosaic composed from the same individual frames --
and is of the order $N_{\rm\sc SM}^{-1}$, where $N_{\rm\sc SM}$
is the number of frames contributing to submosaic SM.
The electron flux at position $(x,y)$ is thus the total flux of all 
frames contributing to the mosaic at that position, weighted by 
the total sky value on each part of the submosaic  
in order to account for frame-to-frame 
differences in transparency and total exposure time. 

We now consider separately individual sources of photometric 
uncertainty within detector regions composed of $N_p$ pixels spread 
over a total area $A$, $N'_p$ of which are unmasked pixels 
that combine to form an unmasked area $A'$.  Throughout, we will assume 
that all portions of the subarea $A$ on the mosaic were constructed from the 
same individual CCD frames.  All expressions for uncertainties are 
expressed in terms of numbers of electrons. 

\subsection{Read Noise: $\sigma_{RN}$}

The noise associated with reading the charge collected in the CCD 
array is associated with every pixel of the array.  For the UT1 test 
camera, this noise has a random distribution with an rms (root-mean-square) 
value of 7.2$e^{-}$.  Over an unmasked area $A'$ on the detector, the 
uncertainty in the flux contributed by read noise is thus

$$
\sigma_{RN}(A') = 7.2 \, {\rm C_{\sc SM}} \, \sqrt{N_f N'_p}~~~,
$$

\noindent
assuming that each of the $N_f$ frames contributing to the mosiac area $A'$ 
had $N'_p$ unmasked pixels.

\subsection{Flat-Fielding: $\sigma_{FF}$}

Flat-fielding was performed by constructing supersky flats from moonless 
UT1 Hubble Deep Field South and EIS images in the same bands 
taken during a 10-day period coinciding with our \gal\ observations.  
Dithering helped to ensure that sky objects did not 
fall on the same portion of the physical detector and could thus 
be removed in the median process (see Sect.~3.1).  Nevertheless, 
Table~\ref{tab:RMSflats} demonstrates that the
fractional rms scatter $\widetilde \sigma _{FF}(A)$
in the flat field  
averaged over different size scales $A$ does not scale with $1/\sqrt{A}$,   
a clear sign that the flat-fielding errors are not purely statistical. 
For scales larger than the gaussian FWHM of the seeing disk
($\sim$1$\arcsec$),
subtle extended light from sky objects may not be entirely removed by the 
supersky flat median process, creating an increase in the flat-fielding 
residuals on these scales.  On the largest scales, systematic errors are 
nearly an order of magnitude larger than those due to counting statistics 
in the flat-fields.

The fractional flat-fielding uncertainties $\widetilde \sigma _{FF}(A')$ 
from Table~\ref{tab:RMSflats} must be multiplied by the total unmasked flux
in area $A'$ on frame $i$, and then combined to yield the flat-fielding 
uncertainty within an area $A'$ on the mosaic.
Since the science frames were dithered by more than 10\arcsec\ in each
direction, larger than 
any area considered here, any $(x,y)$ position on 
the mosaic is constructed with images that were flat-fielded at 
different positions on the physical detector. Thus the flat-fielding 
uncertainties of individual frames can be treated as being 
independent and added in quadrature.  For the mosaic we thus have 

$$
\sigma_{FF}(A') = {\rm C_{\sc SM}} \, \sqrt{
\sum_i^{N_f} 
{ \left( \widetilde \sigma _{FF}(A') \sum_j^{N'_p} e^-_{ij}  
\right) ^2 } }
$$
$$
~~~~~~~~~~~~ = \, 
{\rm C_{\sc SM}} \, 
\widetilde \sigma _{FF}(A') \sqrt{\sum_i^{N_f} \left( \sum_j^{N'_p} e^-_{ij}  
\right) ^2 }~,
$$

\noindent 
where $j$ runs over all unmasked pixels in the area $A'$ on frame $i$ 
and $\widetilde \sigma _{FF}(A')$ can be pulled outside the sum because 
the masking is performed on the mosaic and thus is identical for all 
individual frames $i$.  

\subsection{Photon Noise: $\sigma_{PN}$}

The photon noise is essentially uncorrelated over areas larger than 
the FWHM of the PSF, so that it is given by the square root of the 
number of electrons within that area.  
For a given frame, we thus compute the uncertainty 
due to photon noise within areas $A_{PSF}$ comparable to the PSF,
and then add these in quadrature.  The uncertainties for individual frames 
are independent, and can be added in quadrature to yield the total 
uncertainty due to photon noise within an unmasked area $A'$ on the mosaic. 
Since the uncertainties are proportional to the square root of the 
number of electrons but are then added in quadrature, 
for any area $A' > A_{PSF}$, the resulting photon noise is   

$$
\sigma_{PN}(A') = {\rm C_{\sc SM}} \, \sqrt{
\sum_i^{N_f} 
\sum_k^{N'_p/N_{PSF}} 
\sum_j^{N_{PSF}} e^-_{ijk}  
} 
$$
$$
= \, 
{\rm C_{\sc SM}} \, \sqrt{
\sum_i^{N_f} 
\sum_j^{N'_p} e^-_{ij}  
}
~~,
$$

\noindent
where $N'_p/N_{PSF}$ is the number of PSF-sized areas within $A'$.  
Note that this is nothing more that the square root of the total 
number of electrons recorded by all unmasked pixels from all frames 
contributing to the area $A'$ of the mosaic.  In all of our work, 
we will only quote surface photometry for $A' \geq A_{PSF}$ so that 
all points in our surface brightness profiles are independent, 
and so that the formulation above can be used to calculate photon noise. 

\subsection{Sky Subtraction: $\sigma_{SS}$}

Sky subtraction introduces the same systematic uncertainty to 
every position in the mosaic.   The determination of the sky values 
and their uncertainties $\delta{\cal S}$ 
(${\cal S}_{R} \pm \delta {\cal S}_{R} = {\rm 16651.5 \pm 0.4~e^{-}~pix^{-1}}$, 
${\cal S}_{V} \pm \delta {\cal S}_{V} = {\rm 2950.2 \pm 0.2~e^{-}~pix^{-1}}$)
were discussed in Sect.~\ref{sect:skylevel} and Sect.~\ref{sect:errors}.  
Since the sky values are determined {\it from the mosaic itself\/}, 
the normalization factor ${\rm C_{\sc SM}}$ is already contained in these values. 
We have then simply

$$
\sigma_{SS}(A') = N'_p \, \delta{\cal S} ~~~.
$$

\subsection{Absolute Calibration: $\sigma_{CAL}$}

The absolute calibration, or transformation of our surface brightness 
photometry to a standard system, is not of primary importance to 
many of our scientific results since {\it relative\/} measurements 
from one portion of the mosaic to other portions are more relevant.  
Nevertheless, as explained in Sect.~\ref{sect:photom}, all absolute measurements 
have a {\it fractional\/} uncertainty of 
$\widetilde \sigma _{CAL}(A') \approx 0.05$ 
due to errors in the absolute calibration.   
Thus, for absolute quantities we must also consider 

$$
\sigma_{CAL}(A') = 
\widetilde \sigma _{CAL}(A') \, F_{\rm\sc \, SM}(A')
$$

\noindent 
Again, because the calibration is performed {\it on the 
mosaic itself\/}, the normalization factor ${\rm C_{\sc SM}}$ is 
already contained in these values. 

\subsection{Mosaicing: $\sigma_{M}$}

If the image of \gal\ had been formed without mosaicing, then the total 
normalization constant ${\rm C_{\sc SM}}$ in the first 
equation of the appendix would not be present and thus would introduce 
no uncertainty in the final photometry.  (The uncertainty in ${\rm C_{\sc SM}}$ 
is dominated by the uncertainty in $(\sum^{N_f}_i  {\cal S} _i ) ^{-1}$; 
we ignore here the very much smaller uncertainty in $\zeta$.) 
With mosaicing, {\it relative\/} errors related to the 
uncertainty in the quantity $(\sum^{N_f}_i  {\cal S} _i ) ^{-1}$  
may be introduced between different parts of the mosiac. 
We consider, therefore, the mosaicing uncertainty of photometry 
in submosaic $SM$ relative to the fiducial submosaic $SM_O$ 
(taken to be central submosaic containing flux from all \gal\ frames), to be 
 
$$
\sigma_{M}(A') \equiv \, \sqrt{
(\delta {\rm C_{\sc SM}})^2 +  (\delta {\rm C_{\sc SM_O}})^2 } \, 
\sum_i^{N_f} \sum_j^{N'_p} e^-_{ij}  
$$
$$
= \, \frac{  \sqrt{
 {\rm C^4_{\sc SM}}   \sum_{\rm\sc SM} (\delta {\cal S} _i)^2 + 
 {\rm C^4_{\sc SM_O}} \sum_{\rm\sc SM_O} (\delta {\cal S} _i)^2}}{\zeta}  
\, \sum_i^{N_f} \sum_j^{N'_p} e^-_{ij}
$$

\noindent
where the $(\delta {\cal S} _i)^2$ are the uncertainties in the median sky 
values of individual frames 
summed over those frames contributing to the indicated submosaic. 

\subsection{Surface Brightness Fluctuations: $\sigma_{L}$}

In outlining a method for determining extragalactic distances,
Tonry \& Schneider (1988) derive an expression for the intrinsic
variations in an elliptical galaxy or spiral galaxy bulge.
This fluctuation in surface brightness is due to the counting statistics
of a finite number of unresolved stars contributing flux to each pixel
of a CCD image.

The fluctuations in a single pixel is (\cite{tonry88}):

$$
\sigma_{L}^2 = \overline{g}~t~(\frac{10\,pc}{d})^2~10^{-0.4(\overline{M} - m_1)}
~~,
$$

\noindent
where $\overline{g}$ is the mean number of counts due to the galaxy alone
($\sum_i^{N_f}e^-_{i}-\sum_i^{N_f}{\cal S}_{i}$),
$t$ is the single exposure integration time (seconds), $d$ is the
distance to the source in parsecs, and $m_1$ is the magnitude corresponding
to 1 count per pixel per second in the final image.
This equation, however,
does not take into account the effects of seeing, which strongly reduce
the apparent brightness fluctuations.  
This reduction is significantly greater than 1/$\sqrt{n}$,
where $n$ is the total number of pixels,
for spatial bins larger than the seeing PSF (\cite{morrison94}).
Because of the
large distance to \gal, surface brightness fluctuations make a very
small contribution to our total error budget.
Therefore, rather than simulating the effects of
seeing at various binning scales (as done by \cite{morrison94}),
we compute this error as an upper limit, and simply scale it to our bin
sizes using only a 1/$\sqrt{n}$ factor.

For our data, we take $t$=600 seconds, $d$=102\,Mpc,
$m_1(R)$=21.6\Msqarc\ and
$m_1(V)$=21.3\Msqarc, and $\overline{M}(R)$=-0.7 and 
$\overline{M}(V)$=-0.3 (\cite{tonry90}).  

\subsection{The Total Error Budget}

We are now in a position to combine these different sources of  
uncertainty to arrive at an error budget for our surface brightness 
photometry of \gal. The read noise, flat-fielding, photon noise,
mosaicing and intrinsic surface brightness fluctuation
uncertainties are all independent and statistical, and so can be
added in quadrature, so that

$$
\sigma_{\rm STAT}(A') = \hspace*{6.5cm}
$$
$$
  \sqrt{ \sigma^2_{\rm RN}(A') + \sigma^2_{\rm FF}(A') + \sigma^2_{\rm PN}(A') + 
\sigma^2_{\rm M}(A') + \sigma^2_{\rm L}(A')}~~~.
$$

\noindent
This is the statistical error that will cause random scatter in our 
surface brightness profiles.  
The sky subtraction and absolute calibration 
uncertainties are systematic in that all measurments in the mosaic 
will be affected in the same way by errors in these derived quantities.  
Sky subtraction errors will systematically change the slope of the 
surface brightness profiles; calibration errors will shift the profiles 
by a constant amplitude.  All profiles that we display in
Sect.~\ref{sect:results}
have statistical error bars computed as described in this 
appendix.  The effect of sky subtraction and calibration uncertainties 
must be considered separately.

\end{appendix}

\end{document}